# Transverse spin dynamics of light: the generalized spin-momentum locking for structured guided modes

Peng Shi[1,#], Luping Du[1,#,*], Congcong Li[1], Anatoly V. Zayats[2,†], Xiaocong Yuan[1,‡]

[1]*Nanophotonics Research Centre, Shenzhen Key Laboratory of Micro-Scale Optical Information Technology, Shenzhen University, 518060, China*

[2]*Department of Physics and London Centre for Nanotechnology, King's College London, Strand, London, WC2R 2LS, United Kingdom*

**ABSTRACT:** Spin-momentum locking, a manifestation of topological properties that govern the behavior of surface states, was studied intensively in condensed matter physics resulting in the discovery of topological insulators. The photonic spin-momentum locking was introduced for surface plane-waves which intrinsically carry transverse optical spin, leading to many intriguing phenomena and applications in unidirectional waveguiding, metrology and quantum technologies. In addition to spin, optical waves can exhibit complex properties of vectorial electromagnetic fields, associated with orbital angular momentum or nonuniform intensity variations. Here, by considering both spin and angular momentum, we demonstrate a spin-momentum relationship that governs vectorial properties of guided electromagnetic waves, extending photonic spin-momentum locking to two-dimensional vector field of structured guided waves. This generalized spin-momentum locking was proven both theoretically and experimentally with four kinds of surface structured waves. We further formulate a set of spin-momentum equations which are analogous to the Maxwell's equations. These fundamental equations enable us to understand the underlying physics of optical spin-orbit coupling in guided waves, showing practical importance in spin optics, topological photonics and optical spin-based devices and techniques.

## I. INTRODUCTION

Spin-momentum locking (SML), characterized by unidirectional surface spin states, has given rise to extensive studies in topological insulators [1], superconductor [2], magnon [3], cold-atom systems [4] and Bose–Einstein condensates [5]. The photonic analogy of unidirectional surface spin states was demonstrated with the pseudo-spin by engineering the 'extrinsic' spin-orbit interaction and breaking the time-reversal symmetry in artificial photonic structures with importance for applications [6-8]. On the other hand, due to an 'intrinsic' spin-orbit coupling governed by the Maxwell's field theory, the SML of light was reported and linked to modes with the evanescent fields, such as surface waves or waveguided modes [9-11]. For example, surface plasmon polaritons (SPPs) as surface modes propagating at an insulator/metal interface [12], exhibit features of SML that are analogous to the behavior of surface state on a topological insulator [6-8]. Although photons are bosons with integer spin and surface and waveguided electromagnetic (EM) modes subjected to backscattering [13], in contrast to the helical fermion behavior of surface Dirac modes, they possess the topological $\mathbb{Z}_4$ invariant and hence can transport spin unidirectionally [9]. This intrinsic optical SML lays the foundation for many intriguing phenomena such as spin-controlled unidirectional excitation of surface and waveguided modes, and offers potential applications in photonic integrated circuits, polarisation manipulation, metrology and quantum technologies for generating polarisation entangled states [14-20].

Optical transverse spin [11] plays a key role in the intrinsic SML effect in evanescent waves. In contrast to the conventional longitudinal spin of light with the vector parallel to the propagating direction, the orientation of transverse spin lies in the plane perpendicular to the propagating direction, enabling many intriguing phenomena and applications [21]. The optical transverse spin has been studied extensively in the past few years and been observed in many optical configurations, including evanescent waves [22, 23], waveguided modes [24], interference fields [25], focused beams [26], special structured beams [27], and most recently bulk EM waves in bianisotropic media [28]. These transverse spins in the past were generally defined with respect to the wave-vector **k**, i.e., by calculating the spin angular momentum (SAM) and comparing the spin orientation to the wave-vector. This kind of empirical perspective provide an intuitive way to identify the optical transverse spin in various optical configurations, particularly for the surface plane waves which resulted in the discovery of SML in analogy to electronic topological insulators. This approach however cannot be generalized to a structured surface wave with an arbitrary trajectory. Although one can define a "local" wave vector [29], which is related to the orbital energy flow density (**P₀**), it cannot describe quantitatively optical transverse spin associated with a structured vector wave for which the spin part of the Poynting vector (**Pₛ**) is also important.

Here, we propose a framework to consider the optical transverse spin from the perspective of energy flow density (Poynting vector, **P**). For a scalar field, the photon momentum **p** = $\hbar$**k**, where **k** is the wave vector and $\hbar$ is the reduced Plank constant, determines the energy flow density via $P=c^2 \cdot p$, where $c$ is the speed of light (see Supplementary Text I). Therefore, the wave vector, momentum and energy flow density, which has in general both spin and orbital components ($\mathbf{P} = \mathbf{P_s} + \mathbf{P_o}$), are closely interlinked [29]. The energy flow density can be represented through a current density term in the Hertz potential (see Supplementary Text V). Therefore, the proposed treatment allows also extending the concepts of the dynamics of transverse spin from EM waves to fluid, acoustic and gravitational waves [30-32]. Using the Poynting vector considerations, we derived an intrinsic spin-momentum *curl*-relationship, which, on the one hand, reveals the optical transverse spin dynamics in EM fields and, on the other hand, extends the understanding of the SML from plane evanescent waves to a 2D chiral spin swirl associated with the structured guided modes, therefore, generalizing the optical SML to arbitrary evanescent vector fields. We formulate four equations relating spin and momentum of the electromagnetic wave that are analogous to the Maxwell's equations for EM fields. The results are important for understanding the spin dynamics and spin-orbit coupling in EM waves from RF to UV spectral ranges and for applications in spin optics, topological photonics, polarisation measurements, metrology, the development of robust optical spin-based devices and techniques for quantum technologies.

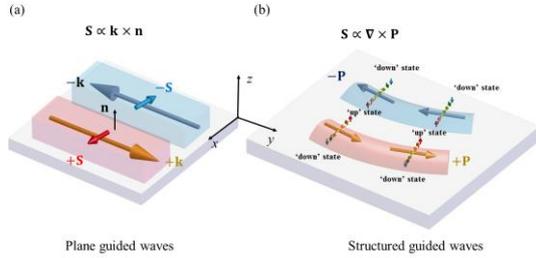

**Figure 1 | Generalization of SML for structured guided modes.** (a) In unstructured, plane guided wave, optical SML results in the transverse spin (**S**) uniformly distributed and parallel to the interface. The spin vector direction is perpendicular to the wave vector **k** and flips if the propagation direction flipped from +**k** to −**k**. (b) In an arbitrary structured guided wave, the optical spin is related to the vorticity of the energy flow density: **P**. The transverse spin vector varies from the 'up' state to the 'down' state around the energy flow, remaining perpendicular to the local wave vector. This forms a chiral swirl of the 2D transverse spin which is locked to the energy propagating direction and fulfills a right handed rule. The direction of the local transverse spin vector flips if the energy flow flipped from forward ('+**P**') to backward ('−**P**').

## II. RESULTS

For an arbitrary EM wave, the *curl* of the energy flow density can be presented as (see Supplementary Text II)

$$\nabla \times \mathbf{P} = c^2 \nabla \times \mathbf{p}$$
$$= \omega^2 \mathbf{S} - \frac{1}{4} \text{Re} \left\{ \begin{array}{l} -(\nabla \otimes \mathbf{E}^*) \cdot \mathbf{H} - (\nabla \otimes \mathbf{E})^T \cdot \mathbf{H}^* \\ +(\nabla \otimes \mathbf{H}^*) \cdot \mathbf{E} + (\nabla \otimes \mathbf{H})^T \cdot \mathbf{E}^* \end{array} \right\}, \quad (1)$$

where **S** is the spin angular momentum, $\omega$ represents the angular frequency of an EM field, **E** and **H** indicate the electric and magnetic field, respectively. Here, the symbol $\otimes$ indicates a dyad vector and * denotes the complex conjugate. The second part on the right-hand side of Eq. (1) has a same structure as the quantum 2-form [33] that generates the Berry phase associated with a circuit, which indicates a spin-orbit interaction in the optical system (see Supplementary Text III for more discussion). In particular, for EM waves with an evanescent field, such as surface or guided waves, an intrinsic spin-momentum relationship can be derived from the Maxwell's theory:

$$\mathbf{S} = \frac{1}{2\omega^2} \nabla \times \mathbf{P} . \quad (2)$$

Since *curl* of a vector field can be regarded as its current vortices, Eq. (2) reveals that the optical spin of an evanescent field is associated with the local vorticity of the EM energy flow and is source-less ($\nabla \cdot \mathbf{S}=0$). The SAM in this case is related to the transverse gradient of the energy flow density. At the same time, the longitudinal optical spin does not fulfill the above spin-momentum relationship. For example, a monochromatic circularly polarized plane-wave bears the SAM aligned parallel to the wave vector, while the *curl* of the Poynting vector vanishes because of the uniformity of the energy flow density over the space. Therefore, the spin-momentum law in Eq. (2) only describes the dynamics of optical transverse spin present in the evanescent waves. It also reveals that, in addition to the optical spin oriented along the surface (in-plane transverse SAM), which has been recently studied intensively, there exists another category of the transverse spin of an evanescent field oriented out of the surface plane. This SAM can be induced by the in-plane energy flow density of the structured guided or surface wave, while the in-plane transverse spin is due to the gradient of energy flow density normal to the interface. The appearance of a transverse spin indicates to the rotation of polarization and hence the phase difference between all the field components of the wave.

The SML in an evanescent plane wave [**Fig. 1(a)**] as demonstrated in previous works [9] is a special, one-dimensional case of SML with the SAM vector aligned along the interface. Assuming the guided mode propagating along $y$-direction and evanescently decaying along $z$-direction, one can deduce the Poynting vector $\mathbf{P} = \hat{\mathbf{y}} \omega \varepsilon / (2\beta) e^{-2k_z z}$ and the SAM $\mathbf{S} = \hat{\mathbf{x}} \varepsilon k_z / (2\omega \beta) e^{-2k_z z}$, where $\varepsilon$ and $\mu$ denote the permittivity and permeability of the medium, respectively, and $\beta$ and $ik_z$ stands for the in-plane and out-of-plane wave vector components, respectively. The energy flow density and SAM of the evanescent plane wave are connected through the generalized spin-momentum relation: $\mathbf{S} = \nabla \times \mathbf{P}/2\omega^2 = -\hat{\mathbf{x}}(\partial P_y/\partial z)/2\omega^2$. For structured evanescent modes with spatially varying intensity distribution, the inhomogeneity of energy flow density can induce several SAM components in different directions. The variation of the energy flow density in $z$-direction induces an in-plane component of the SAM, while its in-plane variations induce a $z$-component. Both are perpendicular to the local energy propagation direction. The relationship between the two components leads to a chiral spin texture with spin vectors swirling around the energy flow [**Fig. 1(b)**]. More importantly, its tendency of directional variation (i.e., the chirality) is locked to the momentum. This is a manifestation of the generalized optical SML associated with an EM evanescent wave.

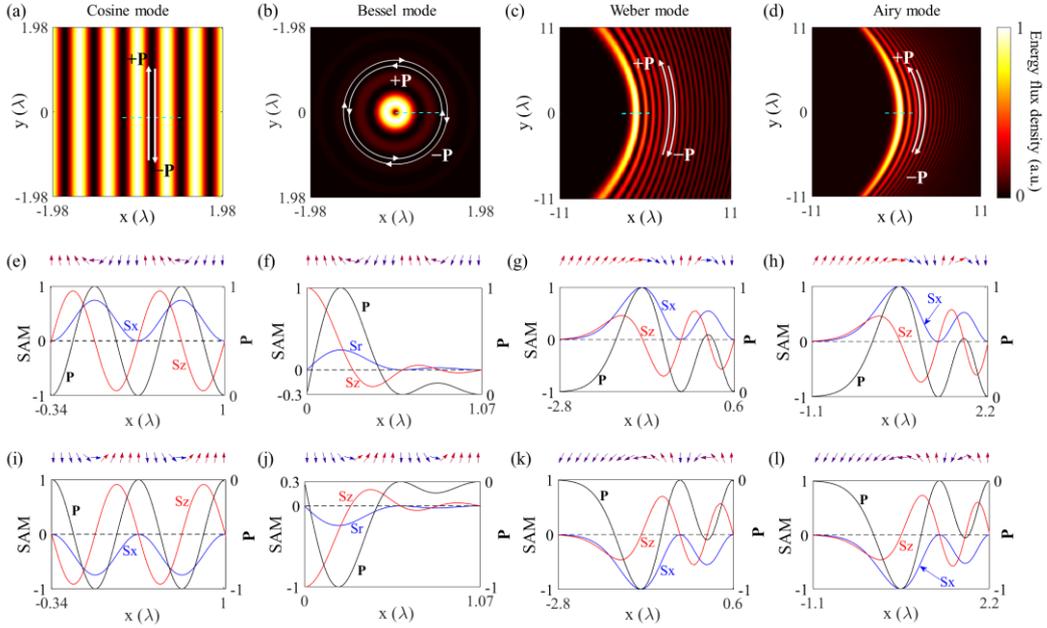

**Figure 2 | Spin-momentum locking in various surface structured waves. (a)-(d),** The spatial distributions of the energy flow density for different structured surface waves: **(a),** surface Cosine beam, **(b),** surface Bessel beam with topological charge $l=\pm 1$; **(c),** surface Weber beam; and **(d),** surface Airy beam. These beams can either propagate in the forward (labelled "+**P**") or backward (labelled "–**P**") directions. **(e)-(h),** Transverse SAM components $S_z$ and $S_x$ and the cross-sections of the energy flow distributions along the green dashed lines in **(a)-(d)** for the beams propagating in direction indicated with the arrow labelled "+**P**". **(i)-(l),** The same as **(e)-(h)** for the beams propagating in the opposite direction indicated with the arrow labelled "–**P**" (c.f. Fig. 1). The inserts at the top of the panels **(e)-(l)** show the local transverse spin vector orientations. The spin vectors are swirling around the energy flow and their local orientations vary from the 'up' to the 'down' states (fulfilling the right handed rule). These orientations are inverted for the waves with the opposite direction of the energy propagating. Note that for the beams with curved trajectory, the spin variation is considered in the plane perpendicular to the local tangential direction of the energy flow. The distance unit is the wavelength of light in vacuum.

It should be noted that the transverse spin discussed here is different from the "spins" in conventional topological photonics, typically called a "pseudo-spin". For a pseudo-spin, the spin-momentum locking is achieved by engineering the spin-orbit interaction in artificial photonic structures in order to break the time-reversal symmetry [8]. In the case of the optical transverse spin of an evanescent wave, the generalized spin-momentum locking is an 'intrinsic' feature of the spin-orbit interaction governed solely by the Maxwell's theory. The nonzero spin Chern number for the structured waves (Supplementary Text III) implies the existence of nontrivial helical states of electromagnetic waves which are strictly locked to the energy propagating direction. However, since the topological $\mathbb{Z}_2$ invariant of these states vanishes owing to the time-reversal symmetry of the Maxwell's equations, there is no protection against (back)scattering. Although the transformation of the two helical states of evanescent waves are not topologically protected against scattering, the SML and the induced unidirectional excitation and propagation are the intrinsic feature of the Maxwell's theory and are topological nontrivial possessing the $\mathbb{Z}_4$ topological invariant. To demonstrate the SML features described by Eq. (2), four types of the EM modes exhibiting evanescent field with inhomogeneous spatial energy distribution were investigated, including the solutions of a wave equation in Cartesian coordinate (Cosine beam) [34], in cylindrical coordinate (Bessel beam) [35], in parabolic coordinate (Weber beam) [36], and in Cartesian coordinate but with a parabolic path (Airy beam) [37] (see Supplementary Text IV for the details of the calculations). The magnitudes of their energy flow densities are shown in **Fig. 2** (top panels), while the beams' propagation directions can either be forward ('+**P**') or backward ('–**P**'). The corresponding cross-section distributions along the dashed lines are shown in the middle and bottom panels for the beams with opposite propagation directions, together with the SAM distributions and the spin vector variation patterns. For all four different types of the beams, the orientation of photon spin vectors varies progressively from the 'up' state to the 'down' state when their photon energies propagate along the forward direction (**Fig. 2**, middle panels). The intrinsic SML present in evanescent waves ensures the topological protection in terms of spin vector swirl being completely determined by the energy flow density. Thus, to observe the reversal of the spin swirling from the 'down' state to the 'up' state, the propagation direction must be reversed (**Fig. 2**, bottom panels). This SML is preserved even for surface modes suffering from the Ohmic losses [12], which influence only the intensity of the wave but not the orientation of photonic spin vector. Note that the spin vector has orientation along the interface at the maxima of the energy flow density, and are normal to it at the nodes. Therefore, a period of spin variation can be defined between the two adjacent nodes of energy flow density which exhibits a similar feature to a topological soliton [38-42].

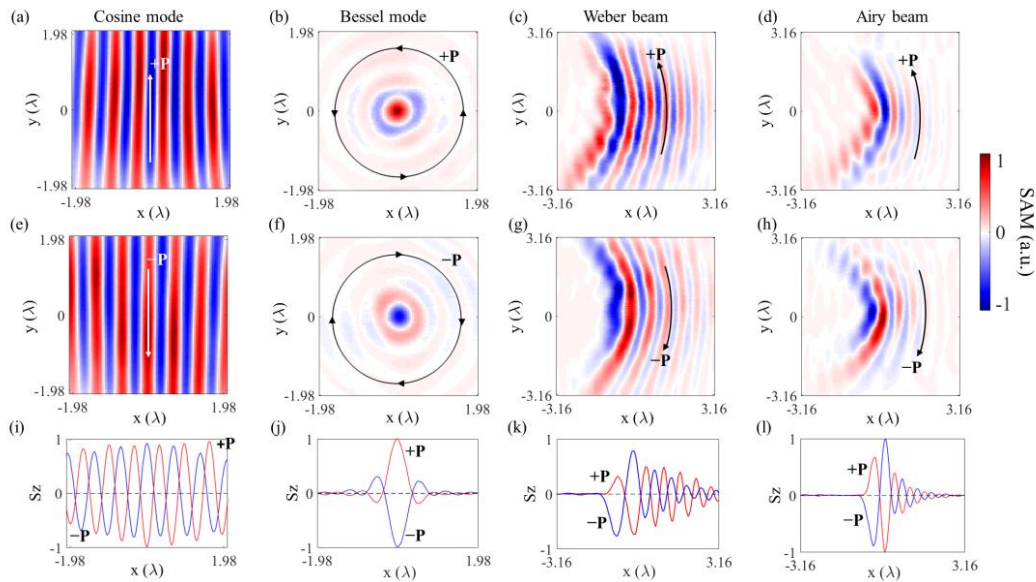

**Figure 3 | Experimental validation of the SML.** The measured out-of-plane SAM components ($S_z$) for **(a), (e), (i)** surface Cosine beam, **(b), (f), (j)** surface Bessel beam, **(c), (g), (k)** surface Weber beam, and **(d), (h), (l)** surface Airy beam: the spatial distributions of $S_z$ spin components for the beams with **(a)-(d)** forward (**+P**) and **(e)-(h)** opposite (**−P**) energy propagating direction, **(i)-(l)** the cross-sections of **(a)-(h)**. The direction of the out-of-plane transverse SAM is inverted for the waves propagating in opposite directions. The distance unit is the wavelength of light in vacuum.

In order to experimentally observe the spin-momentum locking features associated with the structured surface waves and out-of-plane transverse SAM, the experiments were performed on the example of surface plasmon polaritons (SPPs). SPPs were excited under the condition of a total internal reflection using a microscope objective with high numerical aperture NA=1.49. Spatial light modulator and amplitude masks were employed to modulate the phase and wavevector of the excited SPPs to generate the desired plasmonic modes (see Supplementary Text VI for the detailed description of the experiment). A scanning near field optical microscope, which employs a dielectric nanosphere to scatter the SPPs to the far field, and a combination of a quarter waveplate and a polarizer to extract the two circular polarization components of the far-field signal, were used to measure the out-of-plane SAM component $S_z = \varepsilon\beta^2/4\omega k_z^2 (I_{RCP} - I_{LCP})$. The corresponding in-plane spin components were also reconstructed from the measurements (see Supplementary Text VII). The measured distributions of the SAM components are shown in **Fig. 3** and **Figs. S15-S18** for the four types of structured SPP waves propagating in the forward and backward directions. All the predicted SAM and SML features are observed experimentally: (i) the variation of SAM from the positive/negative state to the negative/positive across the beam profile and (ii) the reversal of spin variation when inversing the beam propagation direction.

### III. DISCUSSION

Since the optical energy flow density can be divided into the spin (**P$_s$**) and orbital (**P$_o$**) parts: **P**=**P$_s$**+**P$_o$**, where **P$_s$**=$c^2\nabla\times$**S**/2 and obey the spin-momentum relationship (Eq. 2), we can formulate a set of the Maxwell-like equations linking the transverse spin and the Poynting vector of evanescent EM fields (Table 1). This formulation provides comprehensive understanding of the spin-momentum dynamics in guided waves. The same as variations of **E** field induces **H** field in the Maxwell's equations, equation $\nabla\times$**P** $=2\omega^2$**S** indicates that the spatial variations of the energy flow density induces the transverse spin angular momentum. In the same manner, equation $\nabla\times$**S**=2**P$_s$**/$c^2$=2(**P**-**P$_o$**)/$c^2$ tells us that the spin variation in turn contributes to the energy flux density, with the remainder provided from the orbital part (**P$_o$**) of the Poynting vector. Consolidating spin-momentum equations results in an analogue of a Helmholtz equation $\nabla^2$**S**+$4k^2$**S**= $2\nabla\times$**P$_o$**/$c^2$, which describes spin-orbit interaction in evanescent waves, linking transverse spin and orbital part of the Poynting vector. In both the Helmholtz equation and the last Maxwell's equation, current **J** is an external source of magnetic field; similarly, in the corresponding spin-momentum equations, **P$_o$**, which determines the orbital angular momentum, influences the spin. Since an EM wave in a source-free and homogenous medium can be described with Hertz potential ($\Psi$) satisfying the Helmholtz equation, and the Poynting vector can be calculated from the Hertz potential as $\mathbf{P} \propto (\Psi^*\nabla\Psi - \Psi\nabla\Psi^*)$ [43], one can obtain the spin and orbital properties of the EM guided waves directly from the spin-momentum equations without any knowledge on the electric and magnetic fields (Supplementary Text V).

**Table 1 | Spin/momentum equations and the analogy to the Maxwell's equations.**

| Spin/momentum equations | Maxwell's equations |
|---|---|
| $\nabla \cdot \mathbf{P} = 0$ | $\nabla \cdot \mathbf{E} = 0$ |
| $\nabla \cdot \mathbf{S} = 0$ | $\nabla \cdot \mathbf{H} = 0$ |
| $\nabla \times \mathbf{P} = 2\omega^2 \mathbf{S}$ | $\nabla \times \mathbf{E} = i\omega\mu\mathbf{H}$ |
| $\nabla \times \mathbf{S} = 2(\mathbf{P} - \mathbf{P}_o)/c^2$ | $\nabla \times \mathbf{H} = \mathbf{J} - i\omega\varepsilon\mathbf{E}$ |
| Helmholtz equations | |
| $\nabla^2 \mathbf{S} + 4k^2 \mathbf{S} = 2(\nabla \times \mathbf{P}_o)/c^2$ | $\nabla^2 \mathbf{H} + k^2 \mathbf{H} = -\nabla \times \mathbf{J}$ |

## IV. CONCLUSION

We have demonstrated an intrinsic spin-momentum law which governs the transverse spin dynamics of evanescent EM waves. It was shown that the 1D uniform spin of surface plane wave can be extended to a 2D chiral spin swirl for structured guided modes, providing a manifestation of the generalized photonic spin-momentum locking. Four different types of structured surface waves, including the Cosine beam, Bessel beam, Weber beam and Airy beam, have been investigated both theoretically and experimentally, to demonstrate the concept of the generalized SML. Furthermore, starting from this relation, we obtained a set of spin/momentum related equations that are analogous to the Maxwell's equations. This new optical spin framework can be used to evaluate the spin-orbit coupling in the EM guided waves and for designing specific transverse spin structures, without *a priory* information on the electric and magnetic fields. The generalized intrinsic spin-momentum features could also appear in other types of waves with evanescent field, such as fluid, surface elastic, acoustic and gravitational waves. The effect could be of importance to the development of spin optics for quantum technologies and topological photonics.

## ACKNOWLEDGES

This work was supported, in part, by the National Natural Science Foundation of China grants 61427819, 61490712, 61622504, 61905163 and 61705135, the leadership of Guangdong province program grant 00201505, the Natural Science Foundation of Guangdong Province grant 2016A030312010, the Science and Technology Innovation Commission of Shenzhen grants KQTD2015071016560101, KQTD2017033011044403, KQTD2018041218324255, JCYJ20180507182035270, KQJSCX20170727100838364 and ZDSYS201703031605029, the EPSRC (UK) and the ERC iCOMM project (789340). L.D. acknowledges the support given by the Guangdong Special Support Program. The authors thank Prof. Michael Somekh for proof-reading the manuscript.

## COMPETING INTERESTS

Authors declare no competing interests.

#These authors contributed equally to the work;

**CORRESPONDING AUTHOR:**
*lpdu@szu.edu.cn
†a.zayats@kcl.ac.uk
‡xcyuan@szu.edu.cn

**Transverse spin dynamics of light: the generalized spin-momentum locking for structured guided modes**

Peng Shi[1,#], Luping Du[1,#,*], Congcong Li[1], Anatoly V. Zayats[2,†], Xiaocong Yuan[1,‡]

[1]*Nanophotonics Research Centre, Shenzhen Key Laboratory of Micro-Scale Optical Information Technology, Shenzhen University, 518060, China*

[2]*Department of Physics and London Centre for Nanotechnology, King's College London, Strand, London, WC2R 2LS, United Kingdom*

[#]These authors contributed equally to the work

**Corresponding Author:** [*]lpdu@szu.edu.cn; [†]a.zayats@kcl.ac.uk; [‡]xcyuan@szu.edu.cn

**Contents:**

**I. Spin and orbit decomposition of an arbitrary electromagnetic field**

All the definitions of physical quantities in the manuscript and Supplementary Materials are derived from the Maxwell's theory [s1] and the Riemann–Silberstein (R-S) vector representation for the electromagnetic (EM) fields [s2-s4]. In the following, we only consider the Cartesian coordinates $(x, y, z)$ with directional unit vector $(\hat{\mathbf{x}}, \hat{\mathbf{y}}, \hat{\mathbf{z}})$, while the complex beams in other coordinate systems can also be derived with the same procedure.

The energy flow density (EFD), also known as the Poynting vector (**P**) of an optical wave can be expressed as [s5]:

$$\mathbf{P} = \frac{1}{2}\mathrm{Re}\{\mathbf{E}^* \times \mathbf{H}\}, \tag{S1}$$

where **E** and **H** represent the electric and magnetic fields, respectively. The superscript * denotes the complex conjugate. For a time-harmonic, monochromatic EM wave in the homogeneous medium, the Poynting vector can also be expressed as the momentum density of the field

$$\mathbf{p} = \mathbf{P}/c^2 = \frac{1}{4\omega}\mathrm{Im}\left\{\underbrace{\varepsilon\left[\mathbf{E}^* \times (\nabla \times \mathbf{E})\right]}_{Electric \leftrightarrow \mathbf{p}^e} + \underbrace{\mu\left[\mathbf{H}^* \times (\nabla \times \mathbf{H})\right]}_{magnetic \leftrightarrow \mathbf{p}^m}\right\}. \tag{S2}$$

where $\varepsilon$, $\mu$ denote the permittivity and permeability of the medium, $c$ is the speed of the light and $\omega$ is the angular frequency of wave. The momentum density can be represented as the contributions from the electric ($\mathbf{p}^e$) and magnetic ($\mathbf{p}^m$) field components.

By considering the Gaussian's law of the electric and magnetic fields in a passive and lossless medium, the momentum density of the field can be expressed as [29]

$$\mathbf{p} = \frac{1}{4\omega}\mathrm{Im}\begin{bmatrix} \varepsilon\left(E_x^* \nabla E_x + E_y^* \nabla E_y + E_z^* \nabla E_z\right) \\ + \mu\left(H_x^* \nabla H_x + H_y^* \nabla H_y + H_z^* \nabla H_z\right) \end{bmatrix} + \frac{1}{8\omega}\mathrm{Im}\left[\varepsilon \nabla \times (\mathbf{E}^* \times \mathbf{E}) + \mu \nabla \times (\mathbf{H}^* \times \mathbf{H})\right]. \tag{S3}$$

This expression can be simplified introducing the 6-vector in the homogeneous medium [s2-s4]

$$|\Psi\rangle = \frac{1}{2}\begin{pmatrix} \sqrt{\varepsilon}\mathbf{E} \\ i\sqrt{\mu}\mathbf{H} \end{pmatrix}, \tag{S4}$$

resulting in

$$\mathbf{p} = \frac{1}{\hbar\omega}\langle\Psi|\hat{\mathbf{p}}_3(\mathbf{r})|\Psi\rangle + \frac{1}{2\hbar\omega}i\hat{\mathbf{p}}_3(\mathbf{r}) \times \langle\Psi|\hat{\mathbf{S}}|\Psi\rangle, \tag{S5}$$

where $\hat{\mathbf{p}}_3(\mathbf{r})$ is the Hermitian local momentum operator in the position representation: $\hat{\mathbf{p}}_3(\mathbf{r}) = \left(\delta(\hat{\mathbf{r}} - \mathbf{r})\hat{\mathbf{p}}_3 + \hat{\mathbf{p}}_3\delta(\hat{\mathbf{r}} - \mathbf{r})\right)/2$ with $\hat{\mathbf{p}}_3 = -i\hbar\nabla$ being the momentum operator, $\hat{\mathbf{S}}$ is the spin-1 matrix in SO(3) space [29] and $\hbar$ is the reduced Planck constant. Physically, the first and second terms in Eq. (S5) correspond to the orbital and spin parts of the momentum density, respectively. Therefore, the orbital momentum density (OMD) and spin momentum density (SMD) can be expressed as

$$\mathbf{p}_s = \frac{1}{8\omega}\nabla \times \mathrm{Im}\left[\varepsilon\left(\mathbf{E}^*\times\mathbf{E}\right) + \mu\left(\mathbf{H}^*\times\mathbf{H}\right)\right], \tag{S6a}$$

$$\mathbf{p}_o = \frac{1}{4\omega}\mathrm{Im}\left[\begin{array}{c}\varepsilon\left(E_x^*\nabla E_x + E_y^*\nabla E_y + E_z^*\nabla E_z\right)\\ +\mu\left(H_x^*\nabla H_x + H_y^*\nabla H_y + H_z^*\nabla H_z\right)\end{array}\right], \tag{S6b}$$

respectively. The orbital momentum density $\mathbf{p}_o = \langle\Psi|\hat{\mathbf{p}}_3(\mathbf{r})|\Psi\rangle/\hbar\omega$ is proportional to the local momentum vector. From Eq. (S6a) and the second term in Eq. (S5), the spin angular momentum (SAM) is

$$\mathbf{S} = \frac{1}{4\omega}\mathrm{Im}\left[\varepsilon\left(\mathbf{E}^*\times\mathbf{E}\right) + \mu\left(\mathbf{H}^*\times\mathbf{H}\right)\right]. \tag{S7}$$

## II. Spin-momentum relation for electromagnetic guided waves

### i. General case of the spin-momentum relationship

By employing Eq. (S6b), the *curl* of the OMD can be calculated as

$$\nabla\times\mathbf{p}_o = \frac{1}{4\omega}\mathrm{Im}\left[\begin{array}{c}\varepsilon\left(\nabla E_x^*\times\nabla E_x + \nabla E_y^*\times\nabla E_y + \nabla E_z^*\times\nabla E_z\right)\\ \mu\left(\nabla H_x^*\times\nabla H_x + \nabla H_y^*\times\nabla H_y + \nabla H_z^*\times\nabla H_z\right)\end{array}\right]. \tag{S8}$$

On the other hand, from the relationship between the electric and magnetic fields within the Maxwell's theory, the curl of the SMD can be evaluated as

$$\begin{aligned}\nabla\times\mathbf{p}_s &= -\frac{1}{8\omega}\nabla^2\mathrm{Im}\left[\varepsilon\mathbf{E}^*\times\mathbf{E} + \mu\mathbf{H}^*\times\mathbf{H}\right]\\ &= k^2\mathbf{S} - \frac{1}{4\omega}\mathrm{Im}\left\{\begin{array}{c}\varepsilon\left(\dfrac{\partial\mathbf{E}^*}{\partial x}\times\dfrac{\partial\mathbf{E}}{\partial x} + \dfrac{\partial\mathbf{E}^*}{\partial y}\times\dfrac{\partial\mathbf{E}}{\partial y} + \dfrac{\partial\mathbf{E}^*}{\partial z}\times\dfrac{\partial\mathbf{E}}{\partial z}\right)\\ \mu\left(\dfrac{\partial\mathbf{H}^*}{\partial x}\times\dfrac{\partial\mathbf{H}}{\partial y} + \dfrac{\partial\mathbf{H}^*}{\partial y}\times\dfrac{\partial\mathbf{H}}{\partial y} + \dfrac{\partial\mathbf{H}^*}{\partial z}\times\dfrac{\partial\mathbf{H}}{\partial z}\right)\end{array}\right\},\end{aligned} \tag{S9}$$

where $k=\omega/c$ is the wave vector of the field in the vacuum. By introducing a Dyad's vector given by [s5]

$$\mathbf{r}_1\otimes\mathbf{r}_2 = \begin{pmatrix} x_i^1 x_i^2 & x_j^1 x_i^2 & x_k^1 x_i^2 \\ x_i^1 x_j^2 & x_j^1 x_j^2 & x_k^1 x_j^2 \\ x_i^1 x_k^2 & x_j^1 x_k^2 & x_k^1 x_k^2 \end{pmatrix}, \tag{S10}$$

where the two vectors are $\mathbf{r}_1 = \left(x_i^1, x_i^1, x_i^1\right)^\mathrm{T}$ and $\mathbf{r}_2 = \left(x_i^2, x_i^2, x_i^2\right)^\mathrm{T}$, the *curl* of the momentum density can be expressed as

$$\nabla\times\mathbf{p} = k^2\mathbf{S} - \frac{\varepsilon\mu}{4}\mathrm{Re}\left\{-\left(\nabla\otimes\mathbf{E}^*\right)\cdot\mathbf{H} - \left(\nabla\otimes\mathbf{E}\right)^\mathrm{T}\cdot\mathbf{H}^* + \left(\nabla\otimes\mathbf{H}^*\right)\cdot\mathbf{E} + \left(\nabla\otimes\mathbf{H}\right)^\mathrm{T}\cdot\mathbf{E}^*\right\}. \tag{S11}$$

Accordingly, the *curl* of the energy flow density is

$$\nabla\times\mathbf{P} = v^2\nabla\times\mathbf{p} = \omega^2\mathbf{S} - \frac{\varepsilon\mu v^2}{2}\mathrm{Re}\left\{-\left(\nabla\otimes\mathbf{E}^*\right)_s\cdot\mathbf{H} + \left(\nabla\otimes\mathbf{H}^*\right)_s\cdot\mathbf{E}\right\}, \tag{S12}$$

where $\left(\mathbf{r}_1\otimes\mathbf{r}_2\right)_s = \left\{\left(\mathbf{r}_1\otimes\mathbf{r}_2\right) + \left(\mathbf{r}_1\otimes\mathbf{r}_2\right)^\mathrm{T}\right\}/2$ [s5]. Note that the second part in the right-hand side of Eq. (S12) has a same structure as the quantum 2-form [33] that generates the Berry phase associated with a circuit, which indicates a spin-orbit interaction in the optical system (the relation between this quantum 2-form and Berry phase will be discussed in Section III).

## ii. The spin-momentum relation for the electromagnetic guided waves

The existence of an interface between media with different relative permittivity and permeability breaks the dual symmetry between the electric and magnetic features and the intrinsic connection between the spin and energy flow densities should be considered individually for TM and TE guided modes. We first consider the situation of the guided waves on the example of a transverse magnetic (TM) surface EM wave propagating in *xy*-plane ($H_z = 0$). In this case, the evanescent field exponentially decaying in the *z*-direction can be expressed as $F(x,y)e^{-k_z z}$, where $ik_z$ is the normal to the interface component of the wave vector (Fig. 1). By employing the Maxwell's equations and the Hertz potential theory, the relation between the electric and magnetic field components can be expressed as [s6]

$$E_x = -\frac{k_z}{\beta^2}\frac{\partial E_z}{\partial x} \quad H_x = -\frac{i\omega\varepsilon}{\beta^2}\frac{\partial E_z}{\partial y}$$
$$E_y = -\frac{k_z}{\beta^2}\frac{\partial E_z}{\partial y} \quad H_y = \frac{i\omega\varepsilon}{\beta^2}\frac{\partial E_z}{\partial x} \quad \text{(S13)}$$

where $\beta = \sqrt{k_x^2 + k_y^2}$ represents the in-plane component of the wave vector and is related to $k_z$ by $\beta^2 + (ik_z)^2 = k^2$. After tedious calculations, the *curl* of the energy flow density can be expressed as

$$\nabla \times \mathbf{P} = \omega^2 \mathbf{S} - \frac{1}{4}\text{Re}\begin{pmatrix} i\omega\varepsilon\left(E_y^* E_z - E_y E_z^*\right) \\ i\omega\varepsilon\left(E_z^* E_x - E_z E_x^*\right) \\ i\omega\varepsilon\left(E_x^* E_y - E_x E_y^*\right) + i\omega\mu\left(H_x^* H_y - H_x H_y^*\right) \end{pmatrix} = 2\omega^2 \mathbf{S}. \quad \text{(S14)}$$

Similarly, for the TE evanescent wave ($E_z = 0$), e.g., Bloch surface wave [s7], the field components fulfill the following conditions [s6]:

$$E_x = \frac{i\omega\mu}{\beta^2}\frac{\partial H_z}{\partial y} \quad H_x = -\frac{k_z}{\beta^2}\frac{\partial H_z}{\partial x}$$
$$E_y = -\frac{i\omega\mu}{\beta^2}\frac{\partial H_z}{\partial x} \quad H_y = -\frac{k_z}{\beta^2}\frac{\partial H_z}{\partial y} \quad \text{(S15)}$$

and the curl of the energy flow density also is

$$\nabla \times \mathbf{P} = \omega^2 \mathbf{S} - \frac{1}{4}\text{Re}\left\{i\omega\begin{pmatrix} \mu\left(H_y^* H_z - H_z^* H_y\right) \\ \mu\left(H_z^* H_x - H_x^* H_z\right) \\ \mu\left(H_x^* H_y - H_y^* H_x\right) + \varepsilon\left(E_x^* E_y - E_y^* E_x\right) \end{pmatrix}\right\} = 2\omega^2 \mathbf{S}. \quad \text{(S16)}$$

As the result, for both the TM and TE guided modes, the spin angular momentum and energy flow density (or momentum density) fulfill the relationship

$$\mathbf{S} = \frac{1}{2\omega^2}\nabla \times \mathbf{P} = \frac{1}{2k^2}\nabla \times \mathbf{p}. \quad \text{(S17)}$$

It worth noting that the orbital angular momentum of optical vortex results from the rotation of photon momentum ($\mathbf{L} = \mathbf{r} \times \mathbf{P}_o$, where $\mathbf{r}$ is the position vector and $\mathbf{P}_o$ is the orbital flow density). It is associated to the global phase structure of a beam. Whereas in our case, the transverse optical spin of an EM field originates from the curl/vorticity of energy flow density ($\mathbf{P}_o$).

It is associated to the 'local' rotation of polarization states in the photon transportation.

### III. Discussion on the spin-orbit interaction and spin topological properties for electromagnetic guided waves

**i. Optical Dirac equation and spin-orbit interaction**

The Dirac equation is originally derived for a spin-1/2 particle [s8]:

$$i\hbar \frac{\partial}{\partial t} \Phi = \hat{\mathbf{H}} \Phi = \left( c\boldsymbol{\alpha} \cdot \mathbf{p} + \boldsymbol{\beta} mc^2 \right) \Phi . \tag{S18}$$

Here, $\hat{\mathbf{H}}$ denotes the Hamiltonian operator, $\Phi$ is the electric wave function and the four Dirac matrices can be expressed in terms of the Pauli matrices:

$$\begin{aligned} \boldsymbol{\alpha}_i &= \begin{pmatrix} 0 & \sigma_i \\ \sigma_i & 0 \end{pmatrix} = \sigma_x \odot \sigma_i \\ \boldsymbol{\beta} &= \begin{pmatrix} \sigma_0 & 0 \\ 0 & -\sigma_0 \end{pmatrix} = \sigma_z \odot \sigma_0 \end{aligned}, \tag{S19}$$

where $i = x, y, z$ and the Pauli matrices are

$$\sigma_0 = \begin{pmatrix} 1 & 0 \\ 0 & 1 \end{pmatrix} \quad \sigma_x = \begin{pmatrix} 0 & 1 \\ 1 & 0 \end{pmatrix} \quad \sigma_y = \begin{pmatrix} 0 & -i \\ i & 0 \end{pmatrix} \quad \sigma_z = \begin{pmatrix} 1 & 0 \\ 0 & -1 \end{pmatrix} .$$

For the time-harmonic electromagnetic field propagating in the homogeneous medium, where is in the absence of charges and currents, the Maxwell's equation can be written as

$$\begin{cases} \nabla \cdot \mathbf{E} = 0 \\ \nabla \cdot (\mu \mathbf{H}) = \nabla \cdot (\mu_0 \mu_r \mathbf{H}) = 0 \\ \nabla \times \mathbf{E} = -\mu \frac{\partial \mathbf{H}}{\partial t} = -\mu_0 \frac{\partial \mathbf{H}}{\partial t} - \mu_0 (\mu_r - 1) \frac{\partial \mathbf{H}}{\partial t} \\ \nabla \times \mathbf{H} = \varepsilon \frac{\partial \mathbf{E}}{\partial t} = \varepsilon_0 \frac{\partial \mathbf{E}}{\partial t} + \varepsilon_0 (\varepsilon_r - 1) \frac{\partial \mathbf{E}}{\partial t} \end{cases}, \tag{S20}$$

where $\varepsilon_r$ and $\mu_r$ are the relative permittivity and permeability, respectively. Here, with the identity for arbitrary two vectors $\mathbf{A}$ and $\mathbf{B}$: $\mathbf{A} \times \mathbf{B} = -i(\mathbf{A} \cdot \hat{\mathbf{S}}) \mathbf{B}$ [29], where $\hat{\mathbf{S}}$ is the spin-1 matrix in SO(3) expressed as:

$$\hat{\mathbf{S}} = \{\hat{S}_x, \hat{S}_y, \hat{S}_z\} = \left\{ \begin{pmatrix} 0 & 0 & 0 \\ 0 & 0 & i \\ 0 & -i & 0 \end{pmatrix}, \begin{pmatrix} 0 & 0 & -i \\ 0 & 0 & 0 \\ i & 0 & 0 \end{pmatrix}, \begin{pmatrix} 0 & i & 0 \\ -i & 0 & 0 \\ 0 & 0 & 0 \end{pmatrix} \right\}, \tag{S21}$$

the *curl* operator can be rewritten as

$$\nabla \times = -i(\hat{\mathbf{S}} \cdot \nabla) = \frac{1}{\hbar} \hat{\mathbf{S}} \cdot \hat{\mathbf{p}} . \tag{S22}$$

Firstly, if we consider the electromagnetic field in the homogeneous space, the latter two equations of Eq. (S20) can be written as

$$\hat{\mathbf{H}} |\Psi\rangle = c \begin{pmatrix} \mathbf{0} & \hat{\mathbf{S}} \\ \hat{\mathbf{S}} & \mathbf{0} \end{pmatrix} \cdot \hat{\mathbf{p}} |\Psi\rangle = i\hbar \frac{\partial}{\partial t} |\Psi\rangle . \tag{S23}$$

The Hamiltonian operator $\hat{\mathbf{H}}$ is

$$\hat{\mathbf{H}} = c\hat{\boldsymbol{\tau}} \cdot \hat{\mathbf{p}} = c\begin{pmatrix} \mathbf{0} & \hat{\mathbf{S}} \\ \hat{\mathbf{S}} & \mathbf{0} \end{pmatrix} \cdot \hat{\mathbf{p}}, \tag{S24}$$

where $v\hat{\boldsymbol{\tau}}$ denotes the energy flow density operator (the momentum density operator is $\hat{\boldsymbol{\tau}}/c$) and the corresponding energy flow density can be expressed as

$$\mathbf{P} = c^2 \mathbf{p} = \frac{1}{2}\text{Re}\{\mathbf{E}^* \times \mathbf{H}\} = \left\langle \Psi \left| c\begin{pmatrix} \mathbf{0} & \hat{\mathbf{S}} \\ \hat{\mathbf{S}} & \mathbf{0} \end{pmatrix} \right| \Psi \right\rangle = \langle \Psi | c\hat{\boldsymbol{\tau}} | \Psi \rangle. \tag{S25}$$

Interestingly, the 1st-order partial derivative of position operator is

$$\dot{\mathbf{r}} = \frac{i}{\hbar}\left[\hat{\mathbf{H}}, \mathbf{r}\right] = c\begin{pmatrix} \mathbf{0} & \hat{\mathbf{S}} \\ \hat{\mathbf{S}} & \mathbf{0} \end{pmatrix} = c\hat{\boldsymbol{\tau}}, \tag{S26}$$

which indicates energy flow density operator $v\hat{\boldsymbol{\tau}}$ has the property of a velocity operator and can describe the photon trajectory in accord with the momentum operator. Moreover, the SAM operator of electromagnetic wave is

$$\hat{\boldsymbol{\Sigma}} = \hbar \begin{bmatrix} \hat{\mathbf{S}} & 0 \\ 0 & \hat{\mathbf{S}} \end{bmatrix}, \tag{S27}$$

and the corresponding SAM can be expressed as

$$\mathbf{S} = \frac{1}{4\omega}\text{Im}\{\varepsilon \mathbf{E}^* \times \mathbf{E} + \mu \mathbf{H}^* \times \mathbf{H}\} = \frac{1}{\hbar\omega}\left\langle \Psi \left| \hbar \begin{bmatrix} \hat{\mathbf{S}} & 0 \\ 0 & \hat{\mathbf{S}} \end{bmatrix} \right| \Psi \right\rangle = \frac{1}{\hbar\omega}\langle \Psi | \hat{\boldsymbol{\Sigma}} | \Psi \rangle. \tag{S28}$$

Thus, we can calculate that

$$\dot{\hat{\boldsymbol{\Sigma}}} = \frac{i}{\hbar}\left[\hat{\mathbf{H}}, \hat{\boldsymbol{\Sigma}}\right] = -c\hat{\boldsymbol{\tau}} \times \hat{\mathbf{p}} \quad \text{and} \quad \dot{\hat{\mathbf{L}}} = \frac{i}{\hbar}\left[\hat{\mathbf{H}}, \hat{\mathbf{L}}\right] = c\hat{\boldsymbol{\tau}} \times \hat{\mathbf{p}}, \tag{S29}$$

where the OAM operator is $\hat{\mathbf{L}} = \hat{\mathbf{r}} \times \hat{\mathbf{p}}$. These equations show that the SAM and OAM are not conserved individually, and the evolution of SAM and OAM is relative to $v\hat{\boldsymbol{\tau}} \times \hat{\mathbf{p}}$, which is similar with the *curl* of energy flow density. However, the total angular momentum operator $\hat{\mathbf{J}} = \hat{\mathbf{L}} + \hat{\boldsymbol{\Sigma}}$ is conservative owing to

$$\dot{\hat{\mathbf{J}}} = \frac{i}{\hbar}\left[\hat{\mathbf{H}}, \hat{\mathbf{J}}\right] = \frac{i}{\hbar}\left[\hat{\mathbf{H}}, \hat{\mathbf{L}}\right] + \frac{i}{\hbar}\left[\hat{\mathbf{H}}, \hat{\boldsymbol{\Sigma}}\right] = 0. \tag{S30}$$

The conservative properties of total angular momentum are critical in the analysis of the Chern number and spin-orbit interaction for structured electromagnetic waves.

**ii. Photonic spin Chern number for the structured electromagnetic waves**

The photonic spin Chern number for a plane wave is introduced in Ref. [9]. For a structured light field, the electric/magnetic field can be expressed in a plane-wave basis [s9]

$$\mathbf{E}(\mathbf{r}) = \frac{1}{2\pi} \int_{|\mathbf{k}|=k} \tilde{\mathbf{E}}(\mathbf{k}) e^{i\mathbf{k}\cdot\mathbf{r}} d^2\mathbf{k}$$
$$\mathbf{H}(\mathbf{r}) = \frac{1}{2\pi} \int_{|\mathbf{k}|=k} \tilde{\mathbf{H}}(\mathbf{k}) e^{i\mathbf{k}\cdot\mathbf{r}} d^2\mathbf{k} \quad , \tag{S31}$$

where k = ω/c = |**k**| is the wave number. Transversality constraints for plane waves given by the Maxwell's equations ∇ **E** = ∇ **H** = 0, which result in the requirements $\mathbf{k}\cdot\tilde{\mathbf{E}}(\mathbf{k}) = \mathbf{k}\cdot\tilde{\mathbf{H}}(\mathbf{k}) = 0$ for each single plane wave in the basis, relates the field vector to the wave vector and, therefore, reduces the full 3D vector space of the electromagnetic field components to the 2D subspace of the components tangential to a sphere in the **k** space. Owing to the conservation of total OAM Eq. (S30), this subspace is invariant for the total angular operator $\hat{\mathbf{J}}$, and one can divide it into two parts consistent with the transversality condition [s10]:

$$\hat{\mathbf{J}} = \hat{\mathbf{L}}' + \hat{\mathbf{\Sigma}}', \quad \hat{\mathbf{L}}' = \hat{\mathbf{L}} - \boldsymbol{\kappa} \times (\boldsymbol{\kappa} \times \hat{\mathbf{\Sigma}}) \text{ and } \hat{\mathbf{\Sigma}}' = \boldsymbol{\kappa}(\boldsymbol{\kappa} \cdot \hat{\mathbf{\Sigma}}).$$

where **κ** = **k**/k and the modified OAM and SAM operators $\hat{\mathbf{L}}'$ and $\hat{\mathbf{\Sigma}}'$ can be regarded as projections of the operators $\hat{\mathbf{L}}$ and $\hat{\mathbf{\Sigma}}$ onto the transversality subspace.

Following Ref. [29], we choose an auxiliary vector **e**$_0$ and define two unit vectors as

$$\mathbf{e}_2 = \frac{\mathbf{e}_0 \times \boldsymbol{\kappa}}{|\mathbf{e}_0 \times \mathbf{e}_k|} \quad \mathbf{e}_1 = \mathbf{e}_2 \times \boldsymbol{\kappa}. \tag{S32}$$

The vectors (**e**$_1$, **e**$_2$, **κ**) form a Cartesian frame in which vectors $\tilde{\mathbf{E}}(\mathbf{k})$ and $\tilde{\mathbf{H}}(\mathbf{k})$ lie in the transverse plane (**e**$_1$, **e**$_2$). Next, we introduce the circular polarization basis:

$$\mathbf{e}_+(\mathbf{k}) = \frac{1}{\sqrt{2}}(\mathbf{e}_1 + i\mathbf{e}_2) \quad \mathbf{e}_-(\mathbf{k}) = \frac{1}{\sqrt{2}}(\mathbf{e}_1 - i\mathbf{e}_2), \tag{S33}$$

in which the plane-wave components of the field can be represented as

$$\tilde{\mathbf{E}}(\mathbf{k}) = \mathbf{C}_+(\mathbf{k}) + \mathbf{C}_-(\mathbf{k}) \tag{S34}$$

with **C**$_\sigma$(**k**) = $C_\sigma$(**k**)**e**$_\sigma$(**k**), where σ=±1 and $C_\sigma$(**k**) are the scalar amplitudes of the circularly polarized components. For the structured waves, the OAM of the field is then

$$\mathbf{J} = \int \mathbf{r} \times (\mathbf{p}_s + \mathbf{p}_o) d^2\mathbf{r} = \mathbf{L} + \hat{\mathbf{\Sigma}} = \mathbf{L}' + \hat{\mathbf{\Sigma}}', \tag{S35}$$

where the modulated SAM and OAM components are

$$\hat{\mathbf{\Sigma}}' = \frac{1}{2\omega c} \sum_\sigma \int_{|\mathbf{k}|=k} \sigma \boldsymbol{\kappa} |C_\sigma(\mathbf{k})|^2 d^2\mathbf{k} = \frac{1}{2\omega c} \int_{|\mathbf{k}|=k} \left[ |C_+(\mathbf{k})|^2 - |C_-(\mathbf{k})|^2 \right] \boldsymbol{\kappa} d^2\mathbf{k}$$
$$\mathbf{L}' = \frac{1}{2\omega c} \sum_\sigma \int_{|\mathbf{k}|=k} C_\sigma^*(\mathbf{k}) \cdot \left( -i\mathbf{k} \times \frac{\partial}{\partial \mathbf{k}} - \hat{\mathbf{A}}_B \times \mathbf{k} \right) C_\sigma(\mathbf{k}) d^2\mathbf{k} \tag{S36}$$

It is worth noting that the transformation to the helicity basis is associated with the transition to the local coordinate frame with the z axis attached to the **k**-vector, which induces pure gauge Coriolis-type potential [s11]

$$\hat{\mathbf{A}}_B = -i\hat{U}^\dagger \partial_\mathbf{k} \hat{U}. \tag{S37}$$

This is the Berry gauge field (connection), which corresponds to the monopole curvature $\hat{\mathbf{F}}_B = \partial_{\mathbf{k}} \times \hat{\mathbf{A}}_B = \hat{\sigma} \mathbf{k}/k^3$ with $\hat{\sigma} = \text{diag}(1,-1,0)$ [s12]. By using the Dirac representation and electric–magnetic duality, it can be rewritten as

$$\begin{aligned}\boldsymbol{\Sigma}' &= \langle \Psi(\mathbf{k}) | \hat{\sigma} \boldsymbol{\kappa} | \Psi(\mathbf{k}) \rangle \\ \mathbf{L}' &= \langle \Psi(\mathbf{k}) | \hat{\mathbf{L}} | \Psi(\mathbf{k}) \rangle - \langle \Psi(\mathbf{k}) | \hat{\sigma} \mathbf{A}_B \times \mathbf{k} | \Psi(\mathbf{k}) \rangle \end{aligned}. \tag{S38}$$

From this analysis, the topological Chern numbers for the two helical states can be defined as

$$C^\sigma = \frac{1}{2\pi} \int \langle \Psi(\mathbf{k}) | \hat{\mathbf{F}}_B | \Psi(\mathbf{k}) \rangle d^2\mathbf{k} = \langle \Psi(\mathbf{k}) | 2\sigma | \Psi(\mathbf{k}) \rangle = 2\sigma, \tag{S39}$$

where the normalization condition, which has the meaning of the number of photons in the wave packet, has a form of $N = \langle \Psi(\mathbf{k}) | \Psi(\mathbf{k}) \rangle = 1$ [s10]. Therefore, we can obtain the total Chern number to be

$$C_t = \sum_{\sigma=\pm 1} C^\sigma = 0. \tag{S40}$$

The vanished total Chern number reflects the time-reversal symmetry of non-magnetic Maxwell surface modes. On the other hand, the spin Chern number is

$$C_{spin} = \sum_{\sigma=\pm 1} \sigma C^\sigma = 4. \tag{S41}$$

This nonzero spin Chern number implies that the nontrivial helical states of electromagnetic waves indeed exist and are strictly locked to the energy propagation direction. Despite the existence of such nontrivial helical states at the interface governed by the spin-momentum locking, the topological $\mathbb{Z}_2$ invariant of these states vanishes

$$\upsilon = \frac{C_{spin}}{2} \mod 2 = 0 \tag{S42}$$

owing to the time-symmetry of the Maxwell's equations. Thus, the spin-momentum locking of optical transverse spin discussed here is different from the "pseudo-spin" [6-8] in artificial photonic structures which is engineered to break the time-reversal symmetry, therefore, possessing protection against back-scattering. Although the transformation of the two helical states of evanescent waves are not topologically protected against scattering, the SML and the induced unidirectional excitation and transportation of photons are the intrinsic feature of the Maxwell's theory and are topological nontrivial (possess $\mathbb{Z}_4$ topological invariant).

### iii. Berry phase and spin texture for surface electromagnetic waves

For the general case, the curl of energy flow given in Eq. (S12) can be abbreviated as

$$\nabla \times \mathbf{P} = \omega^2 \mathbf{S} - \frac{\varepsilon \mu \upsilon^2}{2} \text{Re}\left\{-\left(\nabla \otimes \mathbf{E}^*\right)_s \cdot \mathbf{H} + \left(\nabla \otimes \mathbf{H}^*\right)_s \cdot \mathbf{E}\right\}.$$

The second part of this equation has a same structure as the quantum 2-form [33] that generates the Berry phase associated with a circuit, which indicates a spin-orbit interaction in the optical system. In the case of guided modes, this part can be rewritten as

$$\text{Re}\left\{-\left(\nabla \otimes \mathbf{E}^*\right)_s \cdot \mathbf{H} + \left(\nabla \otimes \mathbf{H}^*\right)_s \cdot \mathbf{E}\right\} = 2\omega^2 \mathbf{S} \ . \tag{S43}$$

Thus, the phase change can be obtained as

$$\gamma(C) = \iint_C \text{Re}\left\{-\left(\nabla \otimes \mathbf{E}^*\right)_s \cdot \mathbf{H} + \left(\nabla \otimes \mathbf{H}^*\right)_s \cdot \mathbf{E}\right\} \cdot d\mathbf{a} = \iint_C 2\omega^2 \mathbf{S} \cdot d\mathbf{a} \ , \tag{S44}$$

where $C$ is a two-dimensional connected region of the interface. By applying the general spin-momentum locking equation and the Stokes' theorem to Eq. (S44), the phase change can be rewritten as

$$\gamma(C) = \iint_C 2\omega^2 \mathbf{S} \cdot d\mathbf{a} = \iint \nabla \times \mathbf{P} \cdot d\mathbf{a} = \oint_{\mathbf{a}} \mathbf{P} \cdot d\mathbf{r} \ . \tag{S45}$$

Therefore, the phase change is determined by the optical trajectory around a connected space, which is analogous to the concept of geometric phase in condensed matter physics. In addition, this phase change, determined by the optical trajectory, can also be seen in the optical spin-Hall effect [s13] and optical Magnus effect [s12], which result in the separation of the two helical states and the helicity-depended unidirectional propagation. Thus, the momentum-locked chiral spin texture is also related to the optical-trajectory-determined Berry phase and spin-orbit interaction.

If we consider a region where the energy flow density vanishes at the boundary, the integral also vanishes. Thus, the orientation of SAM should be reversed at the two sides of the extreme point of the energy flow density. This can explain the chiral property of spin texture for the structured guided waves, which is one of the key points of our observations (Fig. 2 of the main text). In addition, the topological number of the chiral spin texture can be $+1/-1$, which denote the spin vector varies from 'up' state to 'down' state or 'down' state to 'up' state accordingly.

Finally, if we consider the integral of the SAM over a total two-dimensional plane, we obtain

$$\iint_{\mathbf{a} \to \infty} \mathbf{S} \cdot d\mathbf{a} = \frac{1}{2\omega^2} \oint_{\mathbf{a} \to \infty} \mathbf{P} \cdot d\mathbf{r} = 0 \tag{S46}$$

owing to the disappearance of energy flow at infinity for a finite size optical beam. Equation (S46) confirms the "local" property of the transverse SAM of surface structured waves [26], in contrast to the longitudinal SAM, for which the integral does not vanish.

## IV. Validation of the spin-momentum relation for various surface waves

We will now verify the above spin-momentum relation for various TM-polarised surface waves. Note that the TE mode evanescent waves can be verified in a same way by exchanging the electric field components with magnetic field components.

**i. Surface plane wave**

For a monochromatic TM mode with an evanescent field decaying in $z$-direction, the $E_z$ field component satisfies the Helmholtz equation:

$$\nabla^2 E_z + k^2 E_z = 0 \ , \tag{S47}$$

where $k = \omega/c$ is the wave vector of the wave. Assuming that the surface wave propagates along y-axis, the electric and magnetic fields can be written as [12]

$$\mathbf{E} = \left(\hat{\mathbf{z}} - i\frac{k_z}{\beta}\hat{\mathbf{y}}\right)\exp[i\beta y - k_z z]$$
$$\mathbf{H} = \frac{\omega\varepsilon}{\beta}\hat{\mathbf{x}}\exp[i\beta y - k_z z]$$
(S48)

where $\hat{\mathbf{x}}, \hat{\mathbf{y}}, \hat{\mathbf{z}}$ are the unit direction vectors. Here, $\beta = k_y$ is the propagation constant. Thus, the energy flow density of the evanescent wave is:

$$\mathbf{P} = \frac{1}{2}\text{Re}\{\mathbf{E}^* \times \mathbf{H}\} = \frac{\omega\varepsilon}{2\beta}\hat{\mathbf{y}}\exp[-2k_z z],$$
(S49)

and the SAM can be calculated to be:

$$\mathbf{S} = \frac{1}{4\omega}\text{Im}\left[\varepsilon\left(\mathbf{E}^* \times \mathbf{E}\right) + \mu\left(\mathbf{H}^* \times \mathbf{H}\right)\right] = \frac{\varepsilon k_z}{2\omega\beta}\hat{\mathbf{x}}\exp[-2k_z z] = -\frac{1}{2\omega^2}\frac{\partial P_y}{\partial z}.$$
(S50)

By examining Eq. (S20) and Eq. (S21), one can find that the spin-momentum relationship for the surface plane wave is satisfied.

ii. **Surface Cosine beam**

The same as above, assuming the beam propagates along y-direction, the z-component of the electric field can be expresses as [34]:

$$E_z = A\cos(k_x x)\exp[ik_y y - k_z z],$$
(S51)

where A is a complex constant.

By employing Eq. (S13), the other field components can be calculated as

$$E_x = A\frac{k_x k_z}{\beta^2}\sin(k_x x)\exp[ik_y y - k_z z] \quad H_x = A\frac{\omega\varepsilon k_y}{\beta^2}\cos(k_x x)\exp[ik_y y - k_z z]$$
$$E_y = -A\frac{ik_y k_z}{\beta^2}\cos(k_x x)\exp[ik_y y - k_z z] \quad H_y = -A\frac{i\omega\varepsilon k_x}{\beta^2}\sin(k_x x)\exp[ik_y y - k_z z]$$
(S52)

The Poynting vector and the *curl* can be calculated as

$$\mathbf{P} = \frac{1}{2}\text{Re}\{\mathbf{E}^* \times \mathbf{H}\} = |A|^2\frac{\omega\varepsilon k_y}{2\beta^2}\hat{\mathbf{y}}\cos^2(k_x x)\exp[-2k_z z]$$
(S53)

and

$$\nabla \times \mathbf{P} = |A|^2\frac{\omega\varepsilon k_y}{2\beta^2}\{\hat{\mathbf{x}}2k_z\cos^2(k_x x) - \hat{\mathbf{z}}k_x\sin(2k_x x)\}\exp[-2k_z z].$$
(S54)

On the other hand, the SAM can be deduced to be:

$$\mathbf{S} = |A|^2\frac{\varepsilon k_y}{4\omega\beta^2}\{\hat{\mathbf{x}}2k_z\cos^2(k_x x) - \hat{\mathbf{z}}k_x\sin(2k_x x)\}\exp[-2k_z z] = \frac{1}{2\omega^2}\nabla \times \mathbf{P},$$
(S55)

which satisfies Eq. (S17). The Poynting vector and SAM distributions for the Cosine beams with forward (+y direction) and backward (–y direction) propagation directions are summarized in Fig. S1, for the special case when $k_x = k_y = \beta\sin(\pi/4)$.

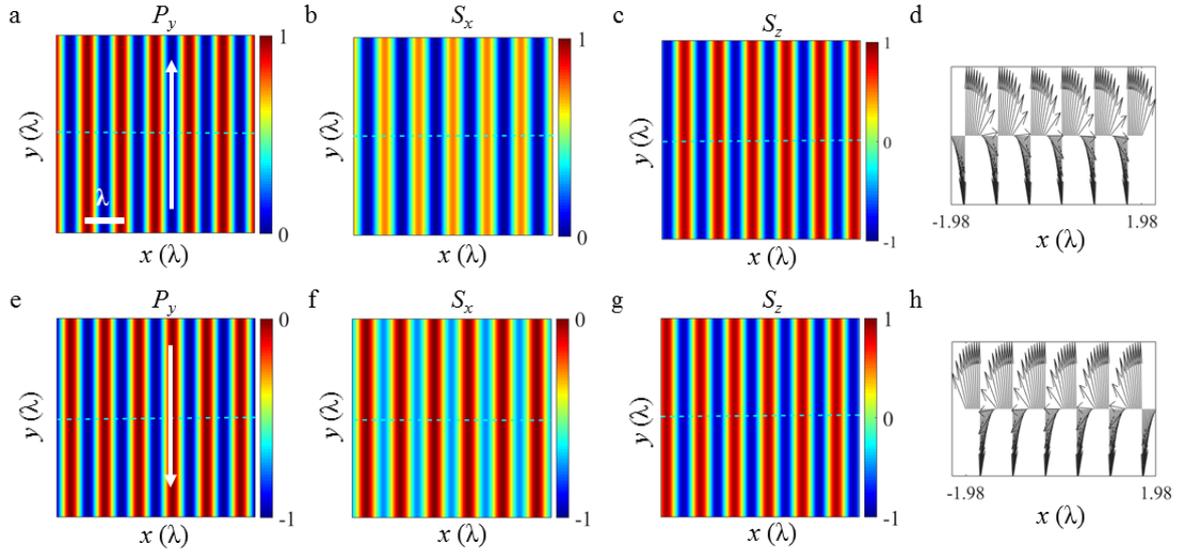

**Fig. S1 | The Poynting vector and spin properties for the surface Cosine beams. a-c,** The energy flow density and SAM distributions for the surface Cosine beam propagating along the +y direction. **d,** The normalized spin vector pattern of the beam along the dashed lines in **a-c**. **e-g,** The energy flow density and SAM distributions for the surface Cosine beam propagating along the -y direction. **h,** the normalized spin vector pattern of the beam along the dashed lines in **e-g**. The arrows in (**a**) and (**e**) show the propagating direction of the beams. The scale bar and the distance unit are the wavelength of light in vacuum.

### iii. Surface Bessel beam

Surface Bessel beams are the solutions of the Maxwell's equations in the cylindrical coordinate $(r, \phi, z)$. The general form of the $z$-component electric field can be expressed as [s14]:

$$E_z = A\beta^2 J_l(\beta r) e^{il\phi} e^{-k_z z}, \tag{S56}$$

where $J_l$ stands for the Bessel function of the first kind with order $l$.

By employing Eq. (S13), the other field components can be calculated as:

$$\begin{aligned}
E_r &= -A k_z \beta J'_l(\beta r) e^{il\phi} e^{-k_z z} & H_r &= A \frac{l\omega\varepsilon}{r} J_l(\beta r) e^{il\phi} e^{-k_z z} \\
E_\phi &= -A \frac{ilk_z}{r} J_l(\beta r) e^{il\phi} e^{-k_z z} & H_\phi &= Ai\omega\varepsilon\beta J'_l(\beta r) e^{il\phi} e^{-k_z z}, \\
E_z &= A\beta^2 J_l(\beta r) e^{il\phi} e^{-k_z z} & H_z &= 0
\end{aligned} \tag{S57}$$

where $J'_l(\beta r)$ stands for the first order derivative of the Bessel function.

The Poynting vector and the *curl* can, therefore, be calculated as

$$\mathbf{P} = \frac{1}{2}\mathrm{Re}\{\mathbf{E}^* \times \mathbf{H}\} = |A|^2 \beta^2 \frac{l\omega\varepsilon}{2r} J_l^2(\beta r) \exp[-2k_z z]\hat{\mathbf{e}}_\phi \tag{S58}$$

and

$$\nabla \times \mathbf{P} = |A|^2 \frac{l\omega\varepsilon\beta^2}{r} \{\hat{\mathbf{e}}_r k_z J_l^2(\beta r) + \hat{\mathbf{e}}_z \beta J_l(\beta r) J_l'(\beta r)\} \exp[-2k_z z]. \quad (S59)$$

Here, $\hat{\mathbf{e}}_r, \hat{\mathbf{e}}_\phi, \hat{\mathbf{e}}_z$ are the unit direction vectors of the cylindrical coordinate system. On the other hand, the SAM can be represented as

$$\mathbf{S} = |A|^2 \frac{l\varepsilon\beta^2}{2\omega r} \{\hat{\mathbf{e}}_r k_z J_l^2(\beta r) + \hat{\mathbf{e}}_z \beta J_l(\beta r) J_l'(\beta r)\} \exp[-2k_z z] = \frac{1}{2\omega^2} \nabla \times \mathbf{P}, \quad (S60)$$

which satisfies Eq. (S17). The Poynting vector and SAM distributions for the surface Bessel beams with topological charge of $l = +/-1$ are given in Fig. S2.

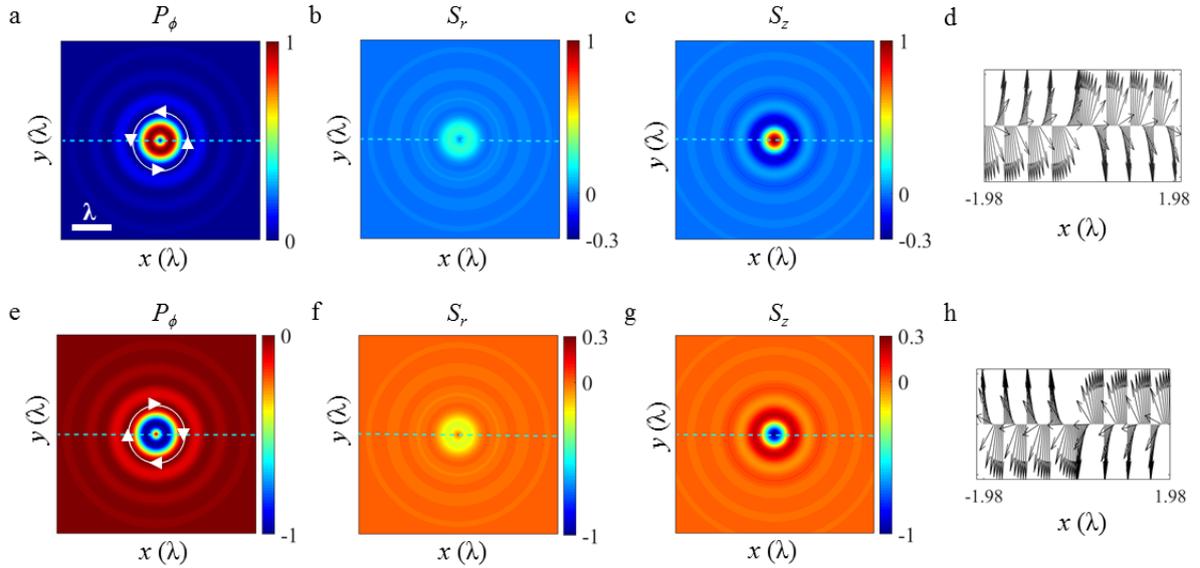

**Fig. S2 | The Poynting vector and spin properties for the surface Bessel beams. a-c,** The energy flow density and SAM distributions for the surface Bessel beam with topological charge of +1, where the energy flow is counter-clockwise (as indicated by arrow in a). **d**, the normalized spin vector pattern of the beam along the dashed lines **a-c**. **e-g**, The energy flow density and SAM distributions for the surface Bessel beam with topological charge of -1, where the energy flow propagates clockwise (as indicated by arrow in e). **h**, the normalized spin vector pattern of the beam along the dashed lines in **e-g**. The scale bar and the distance unit are the wavelength of light in vacuum.

iv.   **Surface Weber beam**

Surface Weber beams are the solutions of the Maxwell's equations in the cylinder parabolic coordinates with a trial solution of $E_z = F(\sigma)G(\tau)e^{-k_z z}$, where $F$ and $G$ are the separation functions of $\sigma$ and $\tau$. In the parabolic coordinates $(\sigma, \tau, z)$, where $\sigma \in (-\infty, \infty); \tau \in [0, \infty); z \in (0, \infty)$, the coordinates are related to those in the Cartesian coordinates by: $x+iy=(\sigma+i\tau)^2/2$ and $z = z$.

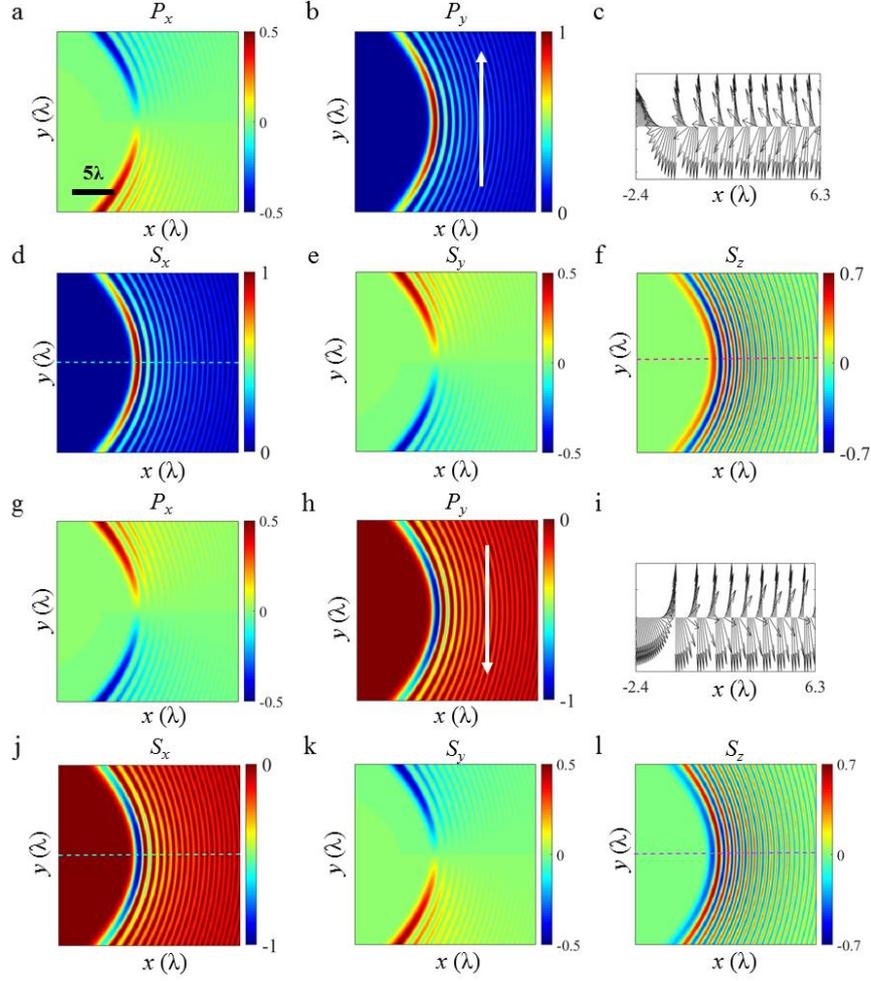

**Fig. S3 | The Poynting vector and spin properties for the surface Weber beams. a-f,** The energy flow density (a,b) and SAM distributions (d-f) for the surface Weber beam propagating along the +y direction. **c**, The normalized spin vector pattern of the beam along the dashed lines in **f**. **g-l,** The energy flow density (g,h) and SAM distributions (j-l) for the surface Weber beam propagating along the -y direction. **i**, the normalized spin vector pattern of the beam along the dashed lines in **l**. The arrows in (**b**) and (**h**) show the propagation direction of the Weber beams. The scale bar and the distance unit are the wavelength of light in vacuum.

By substituting the trial solution into the Helmholtz equation, one can get a characteristic formula for the surface Weber beam as:

$$\frac{1}{F(\sigma)}\left[\frac{\partial^2 F(\sigma)}{\partial \sigma^2}+\sigma^2\beta^2 F(\sigma)\right]+\frac{1}{G(\sigma)}\left[\frac{\partial^2 G(\tau)}{\partial \tau^2}+\tau^2\beta^2 G(\tau)\right]=0. \tag{S61}$$

Since the two terms are functions of independent variables $\sigma$ and $\tau$, they can be separated as:

$$\begin{aligned}\frac{\partial^2 F(\sigma)}{\partial \sigma^2}+(\sigma^2\beta^2+2\beta a)F(\sigma)&=0\\ \frac{\partial^2 G(\tau)}{\partial \tau^2}+(\tau^2\beta^2-2\beta a)G(\tau)&=0\end{aligned}, \tag{S62}$$

where $2\beta a$ is the separation constant. Eq. (S66) demonstrates a two-dimensional propagation-invariant Weber beams in a parabolic cylindrical coordinate [s14].

After solving the equation and transferring back to the Cartesian coordinates and using the relationship: $x = \sigma\tau$, $y = (\sigma^2 - \tau^2)/2$, the general form of the propagating-wave solution is

$$E_z = \frac{1}{\sqrt{2\pi}} \left[ |\Gamma_1|^2 F_e(\sigma) G_e(\tau) + 2i|\Gamma_3|^2 F_o(\sigma) G_o(\tau) \right] e^{-k_z z}$$

$$= \frac{1}{\sqrt{2\pi}} e^{-i\frac{\beta(\sigma^2+\tau^2)}{2} - k_z z} \left\{ \begin{array}{l} |\Gamma_1|^2 \,_2F_1\left[\frac{ia}{2} + \frac{1}{4}; \frac{1}{2}; i\beta\sigma^2\right] \,_2F_1\left[-\frac{ia}{2} + \frac{1}{4}; \frac{1}{2}; i\beta\tau^2\right] \\ +2i|\Gamma_3|^2 \beta\sigma\tau \,_2F_1\left[\frac{ia}{2} + \frac{3}{4}; \frac{3}{2}; i\beta\sigma^2\right] \,_2F_1\left[-\frac{ia}{2} + \frac{3}{4}; \frac{3}{2}; i\beta\tau^2\right] \end{array} \right\}, \tag{S63}$$

where $\Gamma_1 = \Gamma[ia/2 + 1/4]$, $\Gamma_3 = \Gamma[ia/2 + 3/4]$ and $\Gamma[x]$ is the complex Gamma function. Here, $F_1$ is the confluent hypergeometric function of the first kind. The Poynting vector and SAM distributions for the evanescent Weber beams propagating along $+/-y$ directions are illustrated in Fig. S3 for the beam parameter $a=40$. Note that we use the numerical calculation to solve the field components of a Weber beam with Eq. (S13).

**v. Surface Airy wave**

By employing the trial solution of $E_z = A(x, y)e^{ik_y y - k_z z}$, the Helmholtz equation can be simplified to $\frac{\partial^2 \tilde{A}}{\partial x^2} + 2i\beta \frac{\partial \tilde{A}}{\partial y} = 0$, when the small quantity $\frac{\partial^2 \tilde{A}}{\partial y^2}$ representing the variation of the field along the propagating direction can be ignored under the paraxial approximation $\sqrt{\frac{\ln 2}{(aw_m)^2}} \ll \beta$. The solution of the Airy function $\frac{\partial^2 \tilde{A}}{\partial x^2} + 2i\beta \frac{\partial \tilde{A}}{\partial y} = 0$ can be expressed as [s15]

$$\tilde{A}(x, y) = \text{Ai}\left(x_m - \frac{y_m^2}{4} + iay_m\right) e^{i\left[\frac{1}{2}(x_m + a^2)y_m - \frac{1}{12}y_m^3\right]} e^{a\left(x_m - \frac{1}{2}y_m^2\right)}, \tag{S64}$$

where Ai indicates the Airy function of the first kind, $x_m = x/w_m$ and $y_m = y/\beta w_m^2$ are the modulated coordinates, $a$ is a parameter defining the exponential apodization of the Airy beam, and $2w_m$ is the width of the main lobe.

The Poynting vector and SAM distributions for the surface Airy beams propagating along the $+/-y$-direction are summarized in Fig. S4 for the beam parameters $a = 0.01$ and $w_m = 1.1\pi/\beta$.

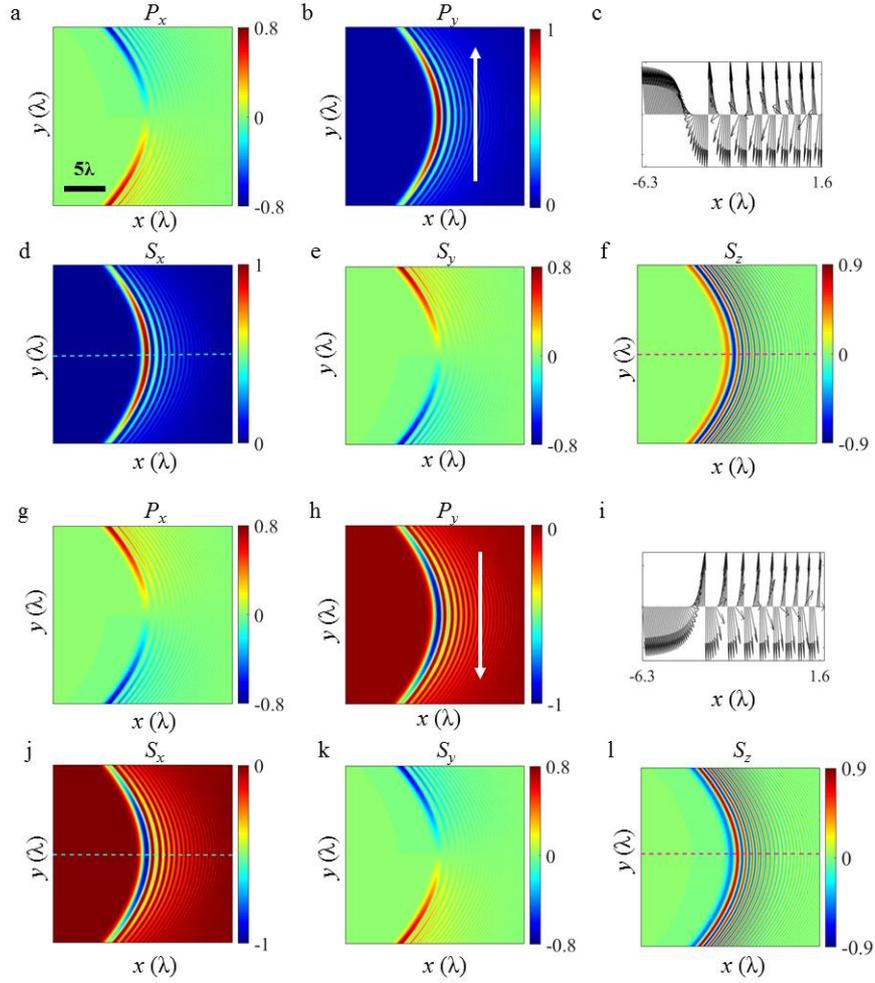

**Fig. S4 | The Poynting vector and spin properties for the surface Airy beams. a-f,** The energy flow density (a,b) and SAM distributions (d-f) for the surface Airy beam propagating in the +y direction. **c,** the normalized spin vector pattern of the beam along the dashed lines in **f**. **g-l,** The energy flow density (g,h) and SAM distributions (j-l) for the surface Airy beam propagating in the −y direction. **i,** the normalized spin vector pattern of the beam along the dashed lines in **l**. The arrows in (**b**) and (**h**) show the propagating direction of the Airy beams. The scale bar and the distance unit are the wavelength of light in vacuum.

### vi. Summary of structured wave solutions

The summary for the above four special types of surface structured beams are shown in **Table. S1**. Note that we use the numerical simulations to solve the field components of the Weber beams with Eq. (S13), as the derivations of the confluent hypergeometric function do not exist. Thus, the analytical expressions of the energy flow density and the SAM for the Weber beams are not provided in Table S1. It was demonstrated that all of the surface structured beams considered above fulfill the generalized spin-momentum relationship.

Nevertheless, there are also many surface mode solutions of the Maxwell's equations which cannot be solved explicitly and, therefore, the spin-momentum relationship cannot be verified analytically. Fortunately, from the classical field theory and quantum field theory [s16], arbitrary EM waves can be expanded by the plane-wave solutions. One can verify that the relationship between the SAM and energy flow density for the superposition of two plane evanescent waves are indeed fulfilled [s17], which indicates that the generalized spin-momentum locking (three-dimensional spin-momentum locking relationship) would be present in an arbitrary structured evanescent field.

**Table. S1. Summarization of $E_z$, EFD and SAM for surface Cosine, Bessel, Weber and Airy beams.**

| Sum | Cosine beam | Bessel beam | Weber beam | Airy beam |
|---|---|---|---|---|
| Helmholtz equation | $\nabla^2 E_z + k^2 E_z = 0$ | | | |
| Trial solution | $A(x,y)e^{-k_z z}$ | $A(r,\phi)e^{-k_z z}$ | $F(\sigma)G(\tau)e^{-k_z \cdot z}$ | $A(x,y)e^{ik_y y - k_z z}$ |
| $E_z$ | $cos(k_x x)e^{ik_y y - k_z z}$ | $k_r^2 J_l(k_r r)e^{il\phi - k_z z}$ | $\frac{1}{\sqrt{2\pi}}[|\Gamma_1|^2 F_e(\sigma)G_e(\tau) + 2i|\Gamma_3|^2 F_o(\sigma)G_o(\tau)]e^{-k_z \cdot z}$ | $\frac{1}{\varepsilon}\widetilde{A}(x,y)e^{ik_y y - k_z z}$ |
| EFDs | $\hat{y}C\cos^2(k_x x)$ | $\hat{\phi}CJ_l^2(k_r r)$ | $\frac{1}{2}\{-Re(E_z^{d*}H_y^d)\hat{x} + Re(E_z^{d*}H_x^d)\hat{y}\}$ | $\left\{\begin{array}{l}\hat{x}C\,Im\left(\widetilde{A}\frac{\partial \widetilde{A}^*}{\partial x}\right)\\ +\hat{y}C\left[Im\left(\widetilde{A}\frac{\partial \widetilde{A}^*}{\partial y}\right) - ik_r|\widetilde{A}|^2\right]\end{array}\right\}$ |
| SAMs | $\frac{C}{2\omega^2}\{-\hat{z}k_x \sin(2k_x x) + \hat{x}2k_z \cos^2(k_x x)\}$ | $\frac{C}{\omega^2}\{\hat{r}k_z J_l^2(k_r r) + \hat{z}k_r J_l(k_r r)J'_l(k_r r)\}$ | $\frac{i}{2\omega}\left\{\begin{array}{l}Im(\varepsilon E_z^{d*}E_y^d)\hat{x}\\ Im(\varepsilon E_z^d E_x^{d*})\hat{y}\\ Im[\varepsilon E_y^{d*}E_x^d + \mu H_x^d H_y^{d*}]\hat{z}\end{array}\right\}$ | $\frac{C}{\omega^2}\left\{\begin{array}{l}\hat{x}k_z\left[Im\left(\widetilde{A}\frac{\partial \widetilde{A}^*}{\partial y}\right) - ik_r|\widetilde{A}|^2\right]\\ -\hat{y}k_z Im\left(\widetilde{A}^*\frac{\partial \widetilde{A}}{\partial x}\right)\\ +\hat{z}\left[Im\left(\frac{\partial \widetilde{A}}{\partial x}\frac{\partial \widetilde{A}^*}{\partial y}\right) + ik_r Re\left(\widetilde{A}\frac{\partial \widetilde{A}^*}{\partial x}\right)\right]\end{array}\right\}$ |
| Parameter | $C = \frac{\omega\varepsilon k_y}{2k_r^2}e^{-2k_z z}$ | $C = \frac{\omega\varepsilon l}{2r}k_r^2 e^{-2k_z z}$ | $y + ix = (\sigma + i\tau)^2/2$ $\sigma \in (-\infty, \infty); \tau \in [0, \infty)$ | $C = \frac{i\omega}{2\varepsilon k_r^2}e^{-2k_z z}$ |

**V. Discussion on spin/momentum locking feature for surface electromagnetic waves**

The demonstrated spin-momentum *curl* relation exhibits the intrinsic locking property between the SAM and optical EFD, and extends the spin-momentum locking to an arbitrary structured guided wave. This is one of the interesting physical effects as demonstrated in the main text. Moreover, starting from this relationship and noting that the Poynting vector of the field can be divided into the spin ($\mathbf{P}_s$) and orbital part ($\mathbf{P}_o$): $\mathbf{P}=\mathbf{P}_s+\mathbf{P}_o$, where $\mathbf{P}_s=c^2\nabla \times \mathbf{S}/2$, we can obtain a set of spin/momentum equations that are analogous to the Maxwell equations (Table. 1 in the main text). As mentioned in the main text, one can obtain the spin and orbital properties of the guided EM waves directly from the spin/momentum equations without any knowledge about the electric and magnetic fields.

In a traditional manner, the spin and orbital angular momentum properties of an EM field is obtained by firstly calculating the electric and magnetic fields. For a time-harmonic monochromatic EM wave in a source free, homogeneous and linear isotropic medium, the Hertz's wave equation independent of the coordinate system is [s6]

$$\nabla^2 \mathbf{\Pi} + k^2 \mathbf{\Pi} = 0. \tag{S65}$$

Eq. (S65) has two types of independent solutions: $\mathbf{\Pi}_e$ and $\mathbf{\Pi}_m$, where the subscript "$e$" and "$m$" denote for the TM and TE mode EM waves, respectively. These result in independent sets of TM waves

$$\begin{aligned} \mathbf{E} &= k^2 \mathbf{\Pi}_e + \nabla(\nabla \cdot \mathbf{\Pi}_e) \\ \mathbf{H} &= -i\omega\varepsilon \nabla \times \mathbf{\Pi}_e \end{aligned}, \tag{S66}$$

and TE waves

$$\begin{aligned} \mathbf{E} &= i\omega\varepsilon \nabla \times \mathbf{\Pi}_m \\ \mathbf{H} &= k^2 \mathbf{\Pi}_m + \nabla(\nabla \cdot \mathbf{\Pi}_m) \end{aligned}. \tag{S67}$$

Assuming the optical axis along the z-direction so that $\mathbf{\Pi}$ can be expressed as $\mathbf{\Pi} = \Psi \hat{\mathbf{z}}$, where $\Psi$ is the model of vector Hertz potential. Obviously, the Hertz potential satisfies the Helmholtz equation:

$$\nabla^2 \Psi + k^2 \Psi = 0. \tag{S68}$$

For the TM waves, the electric and magnetic fields can be calculated as:

$$\begin{aligned} E_x &= -k_z \frac{\partial \Psi}{\partial x} & H_x &= -i\omega\varepsilon \frac{\partial \Psi}{\partial y} \\ E_y &= -k_z \frac{\partial \Psi}{\partial y} & H_y &= i\omega\varepsilon \frac{\partial \Psi}{\partial x} \\ E_z &= \left(k^2 + k_z^2\right)\Psi & H_z &= 0 \end{aligned}. \tag{S69}$$

and for the TE waves, they are:

$$\begin{aligned} E_x &= i\omega\varepsilon \frac{\partial \Psi}{\partial y} & H_x &= -k_z \frac{\partial \Psi}{\partial x} \\ E_y &= -i\omega\varepsilon \frac{\partial \Psi}{\partial x} & H_y &= -k_z \frac{\partial \Psi}{\partial y} \\ E_z &= 0 & H_z &= \left(k^2 + k_z^2\right)\Psi \end{aligned}, \tag{S70}$$

After obtaining the EM field, one can calculate the Poynting vector, spin and orbital angular momentum by the classic definition as in Eq. (S1) and Eq. (S7).

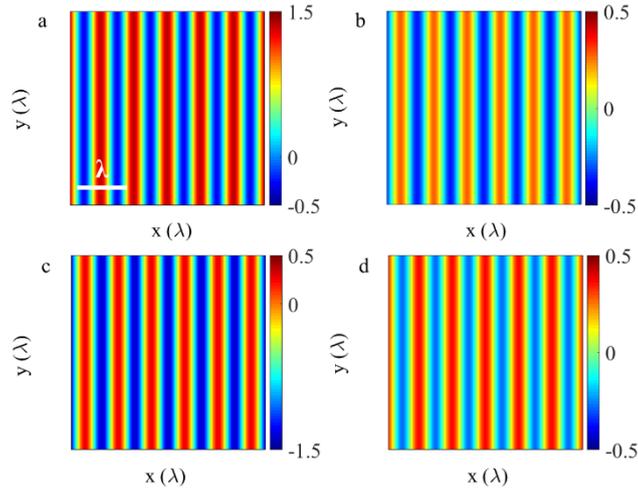

**Fig. S5 | The orbital and spin flow density for the surface Cosine mode. a** and **b** show the $\mathbf{P}_o$ and $\mathbf{P}_s$ for the surface Cosine mode propagating in +y direction with $k_x = \beta \sin(\pi/4)$. **c** and **d** give the $\mathbf{P}_o$ and $\mathbf{P}_s$ for the surface Cosine mode propagating in −y direction with $k_x = \beta \sin(\pi/4)$. All quantities are normalized by the maximum of energy flow density.

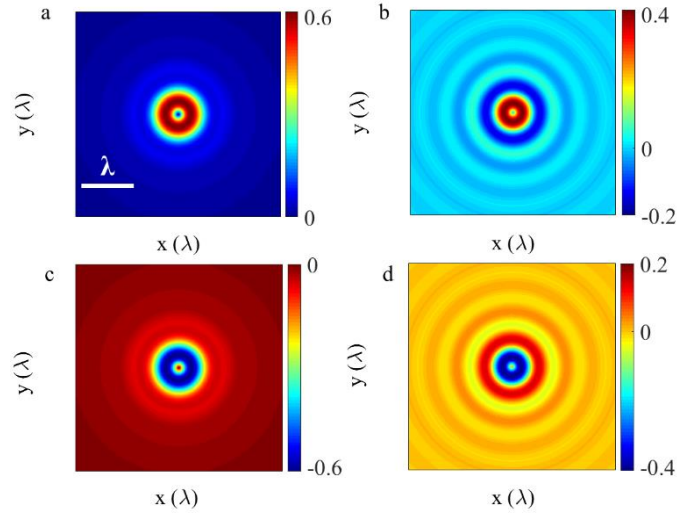

**Fig. S6 | The orbital and spin flow density for the surface Bessel mode. a** and **b** show the $\mathbf{P}_o$ and $\mathbf{P}_s$ for the surface Bessel mode propagating in +φ direction. **c** and **d** indicate the $\mathbf{P}_o$ and $\mathbf{P}_s$ for the surface Bessel mode propagating in −φ direction. All quantities are normalized by the maximum of energy flow density.

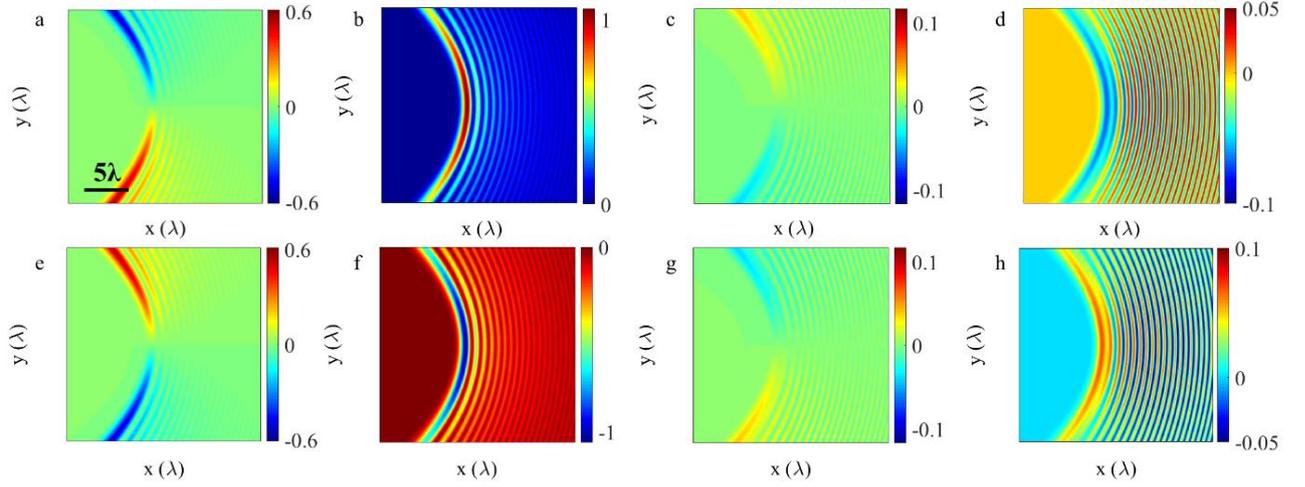

**Fig. S7 | The orbital and spin flow density for the surface Weber mode. a**, **b** show the x and y-components of **P**$_o$ and **c**, **d** exhibit the x and y-components of **P**$_s$ for the surface Weber mode propagating in +y direction with $a = 40$. **e**, **f** show the x and y-components of **P**$_o$ and **g**, **h** exhibit the x and y-components of **P**$_s$ for the surface Weber mode propagating in −y direction with $a = 40$. All quantities are normalized by the maximum of energy flow density.

Noting that the Poynting vector can be calculated from the Hertz potential directly by $\mathbf{P} \propto (\Psi^*\nabla\Psi - \Psi\nabla\Psi^*)$, we can obtain the spin and orbital properties of the guided EM waves directly from the Maxwell's equations without any knowledge about the electric and magnetic fields. It is worth to mention that the classification of the EM field to the TM and TE modes in this case, becomes unnecessary because they have the same spin and orbital properties. In this way, we can obtain the orbital flow density $\mathbf{P}_o = \mathbf{P} + \Delta\mathbf{P}/4k^2$ and spin flow density $\mathbf{P}_s = -\Delta\mathbf{P}/4k^2$, where $\Delta$ is the Laplace operator. The spatial distributions of $\mathbf{P}_o$ and $\mathbf{P}_s$ for the four surface structured waves are shown in Figs. S5-S8, respectively.

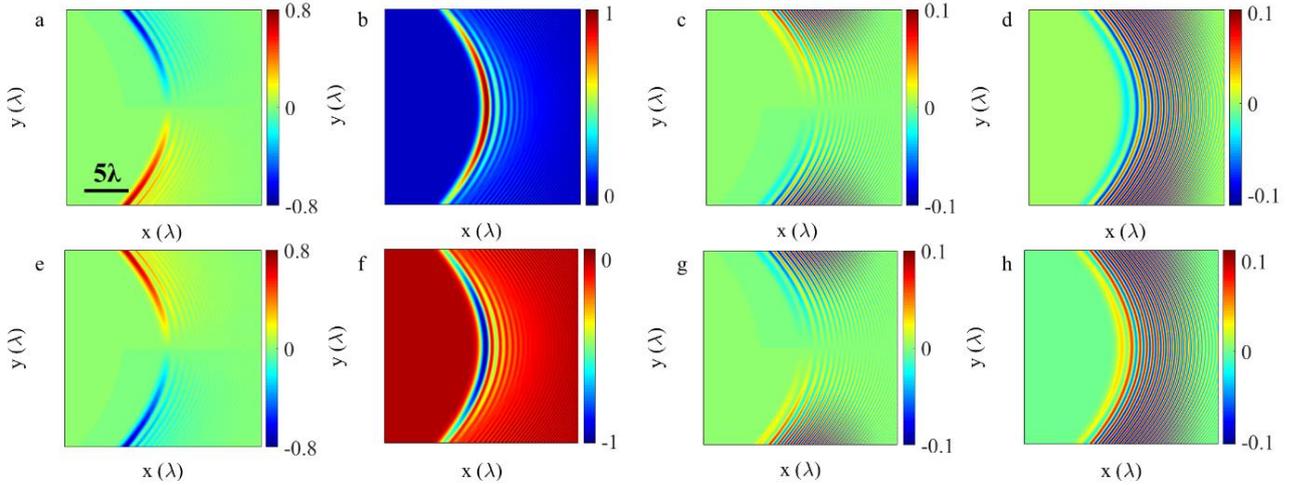

**Fig. S8 | The orbital and spin flow density for the surface Airy mode. a**, **b** show the x and y-components of **P**$_o$ and **c**, **d** exhibit the x and y-components of **P**$_s$ for the surface Airy mode propagating in +y direction with $a = 0.01$. **e**, **f** show the x and y-components of **P**$_o$ and **g**, **h** exhibit the x and y-components of **P**$_s$ for the surface Airy mode propagating in −y direction with $a = 0.01$. All quantities are normalized by the maximum of energy flow density.

## VI. Experimental setup and methods

The experimental setup is shown in **Fig. S9**. The experiment was performed on the example of surface plasmon polaritons (SPPs), which are TM mode evanescent waves supported at a metal- dielectric interface. A He-Ne laser beam with a wavelength of 632.8nm was used as a light source. After a telescope system to expand the beam, a combination of linear polarizer (LP), half-wave plates (HWP), quarter-wave plates (QWPs) and vortex wave plates (VWPs) was employed to modulate the polarization of the laser beam. A spatial light modulator (SLM) was then utilized to modulate the phase of the beam.

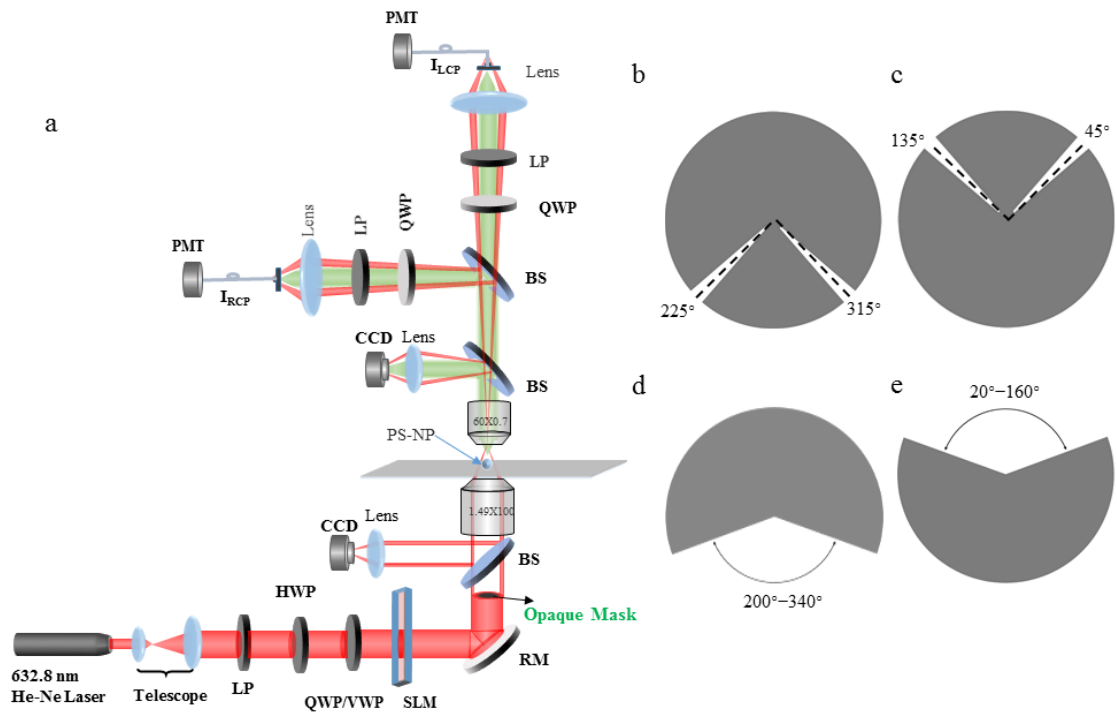

**Fig. S9 | The experimental setup for excitation and mapping of the structured SPP waves**. **a**, Schematic diagram. **b-c**, The designed opaque masks employed in the experiment for generation of a SPP Cosine beams with opposite propagation directions. Incident light was blocked except for the two open segments with the angle spread of 8°. **d**, The designed opaque mask for generation of SPP Weber and Airy beams. Note that a spatial light modulator (SLM) was also employed in order to generate the desired Weber beam and Airy beam. **e**, the mask complementary of (**d**) to generate the corresponding beams with opposite propagation directions to those generated with (d). LP: linear polarizer; HWP: half-wave plate; QWP: quarter-wave plate; VWP: vortex wave plate; RM: reflector mirror; PMT: photo-multiplier tube; BS: non-polarizing beam-splitter.

The structured beam was then tightly focused by an oil-immersion objective (Olympus, NA=1.49, 100×) onto the sample consisting of a thin silver film (45-nm thickness) deposited on a cover slip, to form the desired SPP beams at the air/silver interface. A polystyrene nanosphere of 160 nm radius was immobilized on the silver film surface, as a near-field probe to scatter the SPPs to the far field. The preparation of the sample can be found elsewhere [s18]. The sample was fixed on a Piezo scanning stage (Physik Instrumente, P-545) providing resolution down to 1 nm. A low NA objective (Olympus, NA=0.7, 60×) was employed to collect the scattering radiation from the nanosphere. A combination of quarter wave plate (QWP) and linear polarizer was used to extract the right-handed (RCP) and left-handed (LCP) circular polarization components of the collected signals. Finally, the intensities of RCP and LCP components are measured by a photo-multiplier tube (PMT, Hamamatsu R12829). As the piezo scanning stage raster scanned the near-field region, the distributions of RCP and LCP components can be mapped and used to reconstruct the longitudinal SAM component: $S_z = (I_{RCP} - I_{LCP})\varepsilon\beta^2/4\omega k_z^2$.

**i. Surface Cosine wave**

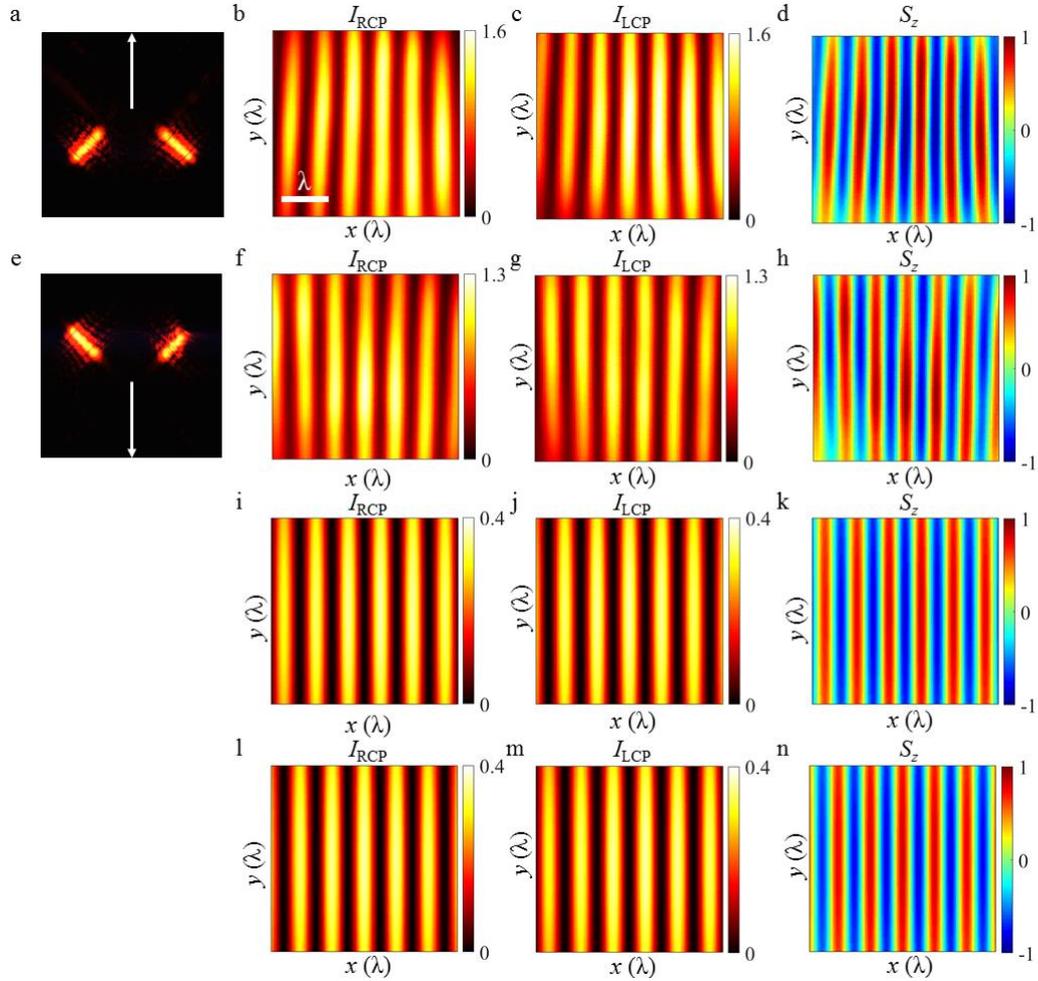

**Fig. S10 | Experimental results for the SPP Cosine beams. a**, The image of the reflected beam captured at the back focal plane of the excitation objective lens when the opaque mask shown in **Fig. S9**(b) was used. In this case, an SPP Cosine beam propagating along the +y direction was generated. **b-d**, The measured intensity distributions of the RCP and LCP components and the retrieved distribution of the longitudinal SAM ($S_z$) component of the SPP Cosine beam. **e-h**, The same as b-d for the opaque mask shown in **Fig. S9**(c) which generates an SPP Cosine beam propagating along the –y direction. The scanning step size in the experiment was 25 nm. The reversal of the optical spin for the two Cosine beams with opposite propagation directions is clearly observed. The arrows in (**a**) and (**e**) show the propagating direction of the generated Cosine beams. **i-k** and **l-n**, The corresponding theoretical calculation results obtained with the vectorial diffraction theory for b-d and e-h, respectively. The distance units is the wavelength of light in vacuum.

To generate the SPP Cosine beams in the experiment, a pair of opaque masks were designed and put right below the objective lens in order to filter the wave vectors in the incident plane of the objective. Incident light was, thus, blocked except for the two open angles with angle spread of 8º (**Fig. S9**(b-c)).

The experimental results are shown in **Fig. S10**(a-h) together with the simulation results in Fig. **S10**(i-n) obtained with the vectorial diffraction theory [s19]. As can be seen, the two opaque masks generate the SPP Cosine beams with opposite propagation directions. The theoretical and experimental results match well and the spin-momentum locking property of the SPP Cosine beams is clearly demonstrated.

## ii. Surface Bessel wave

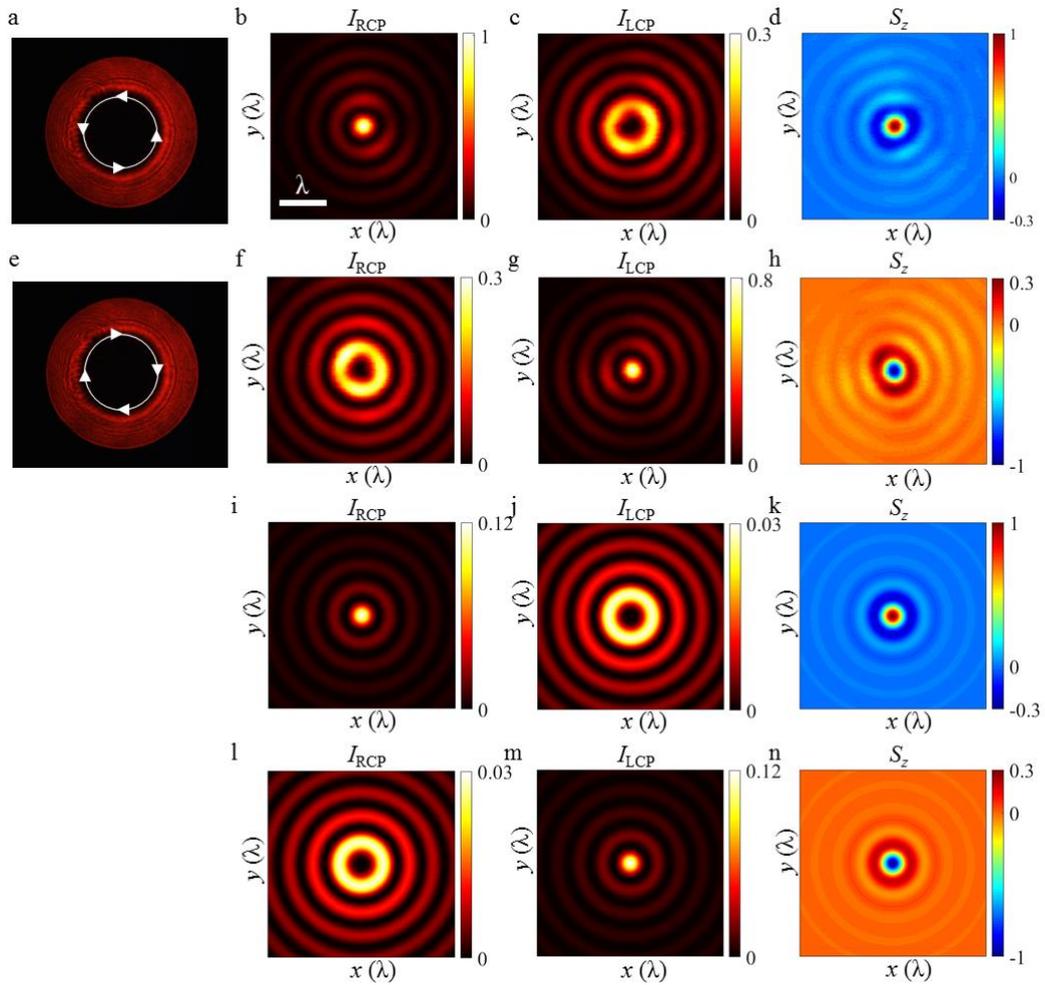

**Fig. S11 | Experimental results for the SPP Bessel beams. a**, The back focal plane image of the reflected beam when LCP light was used to the excite the SPP with topological charge of +1 and the energy propagating counter-clockwise. **b-d**, The mapped $I_{RCP}$ and $I_{LCP}$ distributions and the retrieved distribution of the $S_z$ component of the SPP Bessel beam. **e-h**, The same as b-d for RCP light used to excite the SPP Cosine beam propagating clockwise. The scanning step size in the experiment was 20 nm. It clearly demonstrates the reversal of the optical spin for the two Bessel beams with opposite propagation directions. The arrows in (**a**) and (**e**) indicate the propagating directions of the generated Bessel beams. **i-k** and **l-n**, The corresponding

theoretical calculation results obtained with the vectorial diffraction theory for b-d and e-h, respectively. The distance units is the wavelength of light in vacuum.

The method to generate the SPP Bessel beams can be found elsewhere [35]. Here, we use the left-handed and right-handed CP lights to generate the SPP Bessel beams with topological charge of +1 and –1, respectively. The experimental results are shown in **Fig. S11**, along with the theoretical results for comparison. The spin-momentum locking property for the SPP Bessel beams was clearly demonstrated.

### iii. Surface Weber and Airy beams

The SPP Weber and Airy beams were generated by the vectorial Fourier integral method [s15]. In the experiment, we utilize the opaque masks as shown in **Figs. S9**(d-e) to adjust the spatial frequencies of incident light and employ the SLM to code the phases into the incident beam. The phase for the generation of SPP Weber wave can be expressed as

$$\psi_{Weber} = ia\, k_y/k_r + ia \ln\left( \tan\left( \cos^{-1} \frac{k_y/k_r}{2} \right) \right) \tag{S71}$$

with a parameter $a$=10 was used for the experiment. The phase diagram shown in **Fig. S12**(a), together with the opaque mask shown in **Fig. S9**(d), was employed to generate the SPP Weber beam propagating in +y direction. Similarly, the phase diagram shown in **Fig. S12**(b) and the mask shown in **Fig. S9**(e) were employed to generate the Weber beam propagating in –y direction. The experimental results along with the theoretical simulations are shown in **Fig. S13**.

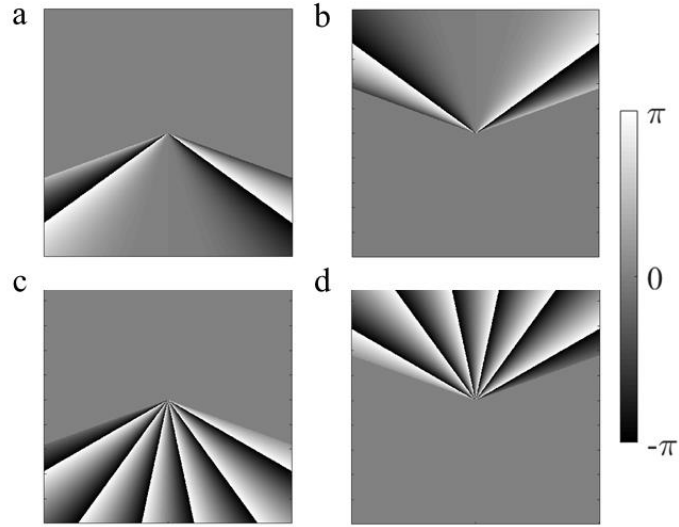

**Fig. S12 | Phase diagrams produced with the SLM for generation of the SPP Weber and Airy beams**. **a-b**, the phases used for generation of the SPP Weber beams with opposite propagating directions: **a** for the +y direction and **b** for the –y direction. **c-d**, the phases for generation of the SPP Airy beams with opposite propagating directions: **c** for the +y direction and **d** for the –y direction.

The method for generating the SPP Airy beams can be found in [s15]. The cubic phase

$$\psi_{Airy} = -i\left\{\left(a_1 \arcsin\left(k_y/\beta\right)\right)^3 \big/ 3 + b_1 \arcsin\left(k_y/\beta\right)\right\} \tag{S72}$$

was encoded into the SLM for generation of the surface Airy beams with parameters $a_1$=0.1 and $b_1$= 15. The phase diagram shown in **Fig. S12**(c), together with the opaque mask shown in **Fig. S9**(d) were employed to generate the SPP Airy beam propagating in +$y$ direction. Similarly, the phase diagram shown in **Fig. S12**(d) and the opaque mask shown in **Fig. S9**(e) was employed to generate the Airy beam propagating in –$y$ direction. The experimental results along with the theoretical simulations are shown in **Fig. S14**.

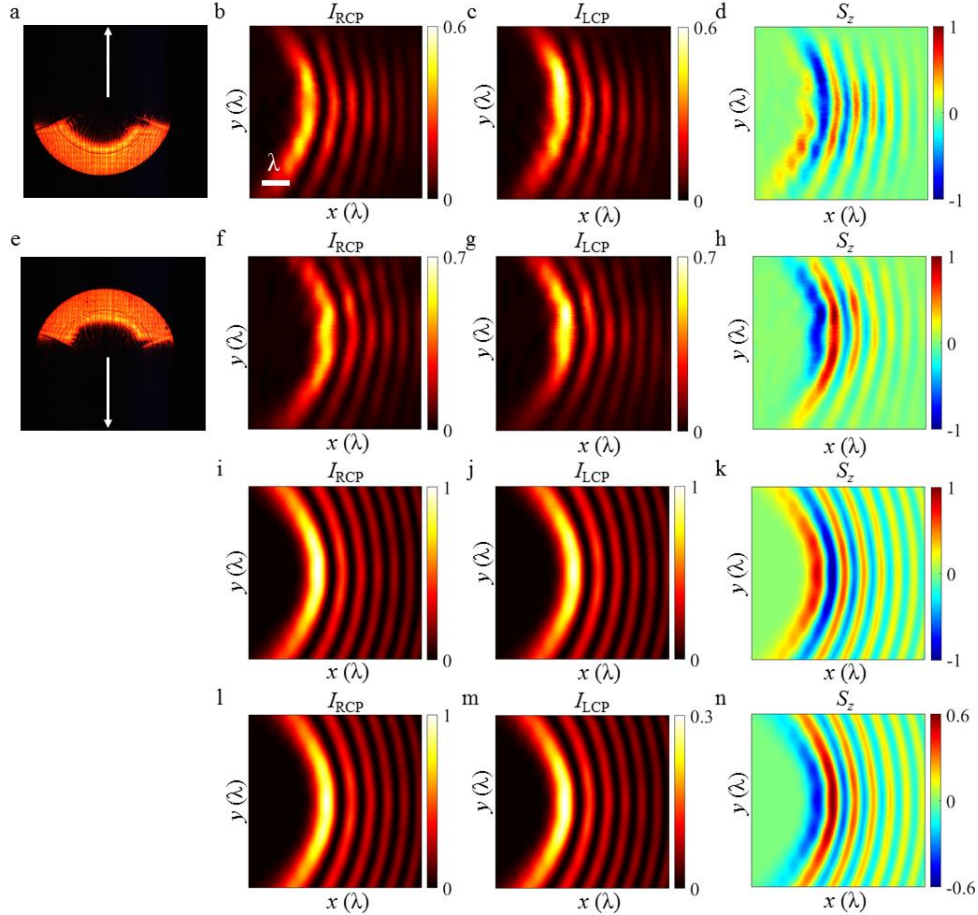

**Fig. S13 | Experimental results for the SPP Weber beams. a**, The back focal plane image of the reflected beam generated with the mask in **Fig. S9**(d) and the phase diagram in **Fig. S12**(a). SPP Weber beam propagates in +y direction. **b-d**, The measured $I_{RCP}$ and $I_{LCP}$ distributions and the retrieved distribution of the $S_z$ component of the SPP Weber beam. **e-h**, The same as **b-d** for the SPP Weber beam propagating along the –y direction (generated with the mask shown in **Fig. S9**(e) the phase diagram shown in **Fig. S12**(b)). The arrows in **a** and **e** indicate the propagating directions of the generated Weber beams. The scanning step size in the experiment was 25 nm. **i-k** and **l-n**, The corresponding theoretical calculation results obtained with the vectorial diffraction theory for **b-d** and **e-h**, respectively. The distance units is the wavelength of light in vacuum.

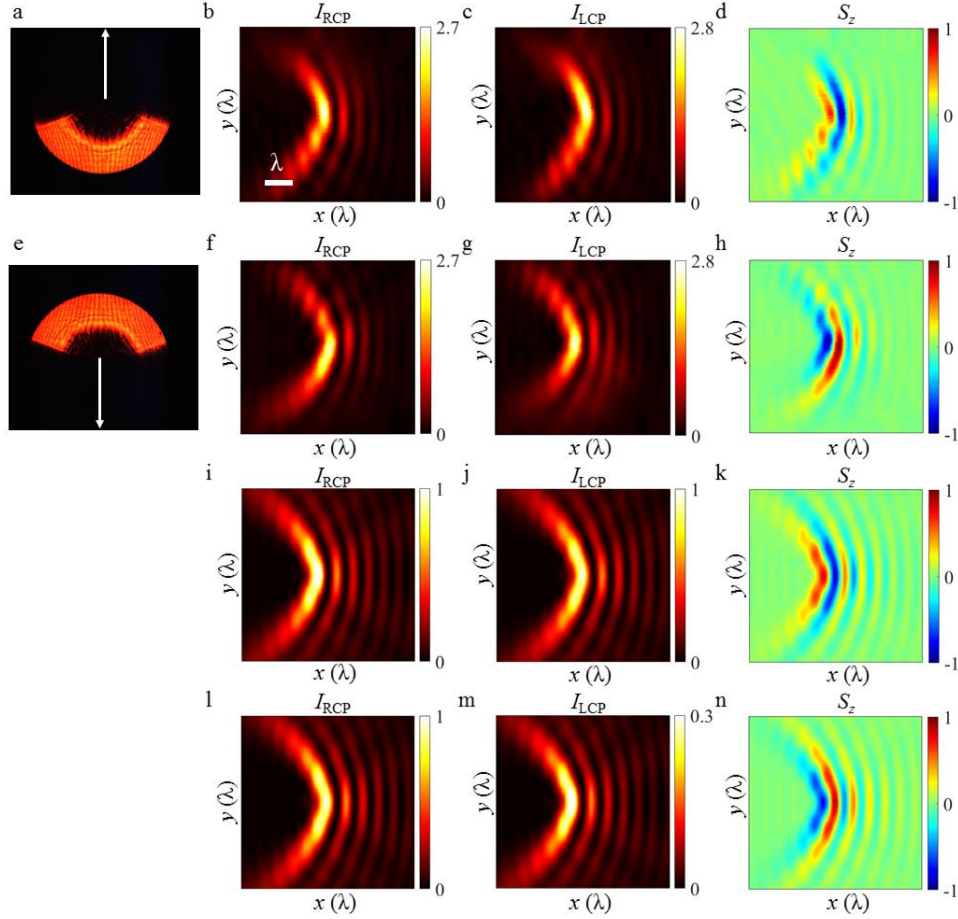

**Fig. S14 | Experimental results for the SPP Airy beams. a**, The back focal plane image of the reflected beam generated with the mask in **Fig. S9**(d) and the phase diagram in **Fig. S12**(c). SPP Airy beam propagates in the +y direction. **b-d**, The measured $I_{RCP}$ and $I_{LCP}$ distributions and the retrieved distribution of the $S_z$ component of the generated SPP Airy beam. **e-h**, The same as **b-d** for an SPP Airy beam propagating in the –y direction (generated with the in **Fig. S9**(e) and the phase diagram in **Fig. S12**(d)). The scanning step size in the experiment was 25 nm. The arrows in **a** and **e** indicate the propagating directions of the generated Airy beams. **i-k** and **l-n**, The corresponding theoretical calculation results obtained with the vectorial diffraction theory for **b-d** and **e-h**, respectively. The distance units is the wavelength of light in vacuum.

## VII. Reconstruction of the in-plane spin angular momentum components

For transverse magnetic (TM) evanescent modes, using the Maxwell's equation, the relation between the transverse electric, magnetic field components and longitudinal electric field component ($E_z$) can be expressed by Eq. (S13). Using the above relationships, the spin angular momentum components can be calculated to be:

$$\begin{cases} S_x = 2Kk_z \, \text{Im}\left( E_z^* \dfrac{\partial E_z}{\partial y} \right) \\ S_y = 2Kk_z \, \text{Im}\left( E_z \dfrac{\partial E_z^*}{\partial x} \right), \\ S_z = 2K \, \text{Im}\left( \dfrac{\partial E_z^*}{\partial x} \dfrac{\partial E_z}{\partial y} \right) \end{cases} \quad \text{(S.73)}$$

where $K = \varepsilon/4\omega\beta^2$ is a physical constant. By carefully examining the expression of spin vectors, we find that

$$\dfrac{\partial S_x}{\partial y} = 2Kk_z \, \text{Im}\left( \dfrac{\partial E_z^*}{\partial y}\dfrac{\partial E_z}{\partial y} + E_z^* \dfrac{\partial^2 E_z}{\partial^2 y} \right) = 2Kk_z \, \text{Im}\left( E_z^* \dfrac{\partial^2 E_z}{\partial^2 y} \right), \quad \text{(S.74a)}$$

and

$$\dfrac{\partial S_y}{\partial x} = 2Kk_z \, \text{Im}\left( \dfrac{\partial E_z}{\partial x}\dfrac{\partial E_z^*}{\partial x} + E_z \dfrac{\partial^2 E_z^*}{\partial x^2} \right) = 2Kk_z \, \text{Im}\left( E_z \dfrac{\partial^2 E_z^*}{\partial x^2} \right). \quad \text{(S.74b)}$$

On the other hand, the $z$-component electric field $E_z$ satisfies Helmholtz equation:

$$\nabla^2 E_z + k^2 E_z = \dfrac{\partial^2 E_z}{\partial x^2} + \dfrac{\partial^2 E_z}{\partial y^2} + \beta^2 E_z = 0. \quad \text{(S.75)}$$

Therefore, by employing the Eq. (S75), the Eq. (S74a) can be converted into

$$\dfrac{\partial S_x}{\partial y} = 2Kk_z \, \text{Im}\left( -E_z^* \dfrac{\partial^2 E_z}{\partial x^2} - \beta^2 E_z^* E_z \right) = 2Kk_z \, \text{Im}\left( E_z \dfrac{\partial^2 E_z^*}{\partial x^2} \right) = \dfrac{\partial S_y}{\partial x}. \quad \text{(S.76)}$$

By employing the conservation law of the spin vectors ($\nabla \cdot \mathbf{S} = 0$ which can be deduced from Eq. (S.7), one can obtain a pair of linearly partial differential equations for transverse spin vectors:

$$\dfrac{\partial^2 S_x}{\partial x^2} + \dfrac{\partial^2 S_x}{\partial y^2} = 2k_z \dfrac{\partial S_z}{\partial x}, \quad \text{(S.77a)}$$

$$\dfrac{\partial^2 S_y}{\partial x^2} + \dfrac{\partial^2 S_y}{\partial y^2} = 2k_z \dfrac{\partial S_z}{\partial y}. \quad \text{(S.77b)}$$

The transverse SAM component $S_x$ can be expressed by the Fourier expansion

$$S_x = \sum_{n=-\infty}^{\infty} \sum_{m=-\infty}^{\infty} \left( A_m \sin\dfrac{m\pi x}{L_x} + B_m \cos\dfrac{m\pi x}{L_x} \right)\left( C_n \sin\dfrac{n\pi y}{L_y} + D_n \cos\dfrac{n\pi y}{L_y} \right), \quad \text{(S.78)}$$

where $L_x$ and $L_y$ are boundary condition parameters. $A_m$, $B_m$, $C_m$ and $D_m$ are constants to be determined. From Eq. (S.77a) and Eq. (S.78), we can get that:

$$-\sum_{n=-\infty}^{\infty} \sum_{m=-\infty}^{\infty} \lambda_{mn}\left( A_m \sin\dfrac{m\pi x}{L_x} + B_m \cos\dfrac{m\pi x}{L_x} \right)\left( C_n \sin\dfrac{n\pi y}{L_y} + D_n \cos\dfrac{n\pi y}{L_y} \right) = 2k_z \dfrac{\partial S_z}{\partial x}, \quad \text{(S.79)}$$

where $\lambda_{mn} = (\dfrac{m\pi}{L_x})^2 + (\dfrac{n\pi}{L_y})^2$ is the eigenvalue. The expansion coefficients can be determined through the Fourier integral of equation (S.79). Omitting the complex mathematical process, we obtain that $S_x$ can be solved to be:

$$\begin{cases} S_x = \sum_{n=-\infty}^{\infty} \sum_{m=-\infty}^{\infty} \begin{pmatrix} A'_m \sin\dfrac{m\pi x}{L_x} \sin\dfrac{n\pi y}{L_y} + B'_m \sin\dfrac{m\pi x}{L_x} \cos\dfrac{n\pi y}{L_y} \\ +C'_m \cos\dfrac{m\pi x}{L_x} \sin\dfrac{n\pi y}{L_y} + D'_m \cos\dfrac{m\pi x}{L_x} \cos\dfrac{n\pi y}{L_y} \end{pmatrix} \\ A'_m = -\dfrac{1}{4L_x L_y \lambda_{mn}} \int_0^{L_x} \int_0^{L_y} 2k_z \dfrac{\partial S_z}{\partial x} \sin\dfrac{m\pi x}{L_x} \sin\dfrac{n\pi y}{L_y} dxdy \\ B'_m = -\dfrac{1}{4L_x L_y \lambda_{mn}} \int_0^{L_x} \int_0^{L_y} 2k_z \dfrac{\partial S_z}{\partial x} \sin\dfrac{m\pi x}{L_x} \cos\dfrac{n\pi y}{L_y} dxdy \\ C'_m = -\dfrac{1}{4L_x L_y \lambda_{mn}} \int_0^{L_x} \int_0^{L_y} 2k_z \dfrac{\partial S_z}{\partial x} \cos\dfrac{m\pi x}{L_x} \sin\dfrac{n\pi y}{L_y} dxdy \\ D'_m = -\dfrac{1}{4L_x L_y \lambda_{mn}} \int_0^{L_x} \int_0^{L_y} 2k_z \dfrac{\partial S_z}{\partial x} \cos\dfrac{m\pi x}{L_x} \cos\dfrac{n\pi y}{L_y} dxdy \end{cases}. \quad (S.80)$$

In a similar manner, from Eq. (S.77b), $S_y$ can be expressed as:

$$\begin{cases} S_y = \sum_{n=-\infty}^{\infty} \sum_{m=-\infty}^{\infty} \begin{pmatrix} A''_m \sin\dfrac{m\pi x}{L_x} \sin\dfrac{n\pi y}{L_y} + B''_m \sin\dfrac{m\pi x}{L_x} \cos\dfrac{n\pi y}{L_y} \\ +C''_m \cos\dfrac{m\pi x}{L_x} \sin\dfrac{n\pi y}{L_y} + D''_m \cos\dfrac{m\pi x}{L_x} \cos\dfrac{n\pi y}{L_y} \end{pmatrix} \\ A''_m = -\dfrac{1}{4L_x L_y \lambda_{mn}} \int_0^{L_x} \int_0^{L_y} 2k_z \dfrac{\partial S_z}{\partial y} \sin\dfrac{m\pi x}{L_x} \sin\dfrac{n\pi y}{L_y} dxdy \\ B''_m = -\dfrac{1}{4L_x L_y \lambda_{mn}} \int_0^{L_x} \int_0^{L_y} 2k_z \dfrac{\partial S_z}{\partial y} \sin\dfrac{m\pi x}{L_x} \cos\dfrac{n\pi y}{L_y} dxdy \\ C''_m = -\dfrac{1}{4L_x L_y \lambda_{mn}} \int_0^{L_x} \int_0^{L_y} 2k_z \dfrac{\partial S_z}{\partial y} \cos\dfrac{m\pi x}{L_x} \sin\dfrac{n\pi y}{L_y} dxdy \\ D''_m = -\dfrac{1}{4L_x L_y \lambda_{mn}} \int_0^{L_x} \int_0^{L_y} 2k_z \dfrac{\partial S_z}{\partial y} \cos\dfrac{m\pi x}{L_x} \cos\dfrac{n\pi y}{L_y} dxdy \end{cases}. \quad (S.81)$$

Through the measured longitudinal SAM component $S_z$, the transverse SAM components can be obtained from Eq. (S.80) and Eq. (S.81), and a complete photonic spin vector can be constructed.

In addition, we can employ the symmetry of surface electromagnetic modes to simplify the calculation further. Taking the surface Bessel mode for example, $S_z$ is mirror symmetric with respect to $x$, $y$-axes, which can be expressed as $\hat{m}_x S_z = S_z$ and $\hat{m}_y S_z = S_z$, where $\hat{m}_i$ is the mirror operator with respect to $i$-axis in Cartesian coordinates. Noted that the $i$-coordinate is mirror antisymmetric with respect to $i$-axis is mirror symmetric with respect to another axis, so that we have $\hat{m}_x x = -x$ and $\hat{m}_y x = x$. Thus, we can get the following relationships:

$$\hat{m}_x \frac{\partial S_z}{\partial x} = -\frac{\partial S_z}{\partial x} \quad \hat{m}_y \frac{\partial S_z}{\partial x} = \frac{\partial S_z}{\partial x}, \quad (S.82)$$

which can be utilized to simplify the Eqs. (S.80) and (S.81):

$$\begin{cases} S_x = \sum_{n=-\infty}^{\infty} \sum_{m=-\infty}^{\infty} \left( A_m \sin\dfrac{m\pi x}{L_x} \cos\dfrac{n\pi y}{L_y} \right) \\ A_m = -\dfrac{1}{4L_x L_y \lambda_{mn}} \int_0^{L_x} \int_0^{L_y} 2k_z \dfrac{\partial S_z}{\partial x} \sin\dfrac{m\pi x}{L_x} \cos\dfrac{n\pi y}{L_y} dxdy \end{cases}, \quad (S.83a)$$

$$\begin{cases} S_y = \sum_{n=-\infty}^{\infty} \sum_{m=-\infty}^{\infty} \left( B_m \cos\frac{m\pi x}{L_x} \sin\frac{n\pi y}{L_y} \right) \\ B_m = -\frac{1}{4L_x L_y \lambda_{mn}} \int_0^{L_x} \int_0^{L_y} 2k_z \frac{\partial S_z}{\partial y} \cos\frac{m\pi x}{L_x} \sin\frac{n\pi y}{L_y} dxdy \end{cases} \quad (S.83b)$$

Through this method, we reconstructed from the experimental measurements the in-plane SAM components for the four structured waves as shown in **Figs. S15-18**, where the results obtained from the experiment match well with the theoretical simulations.

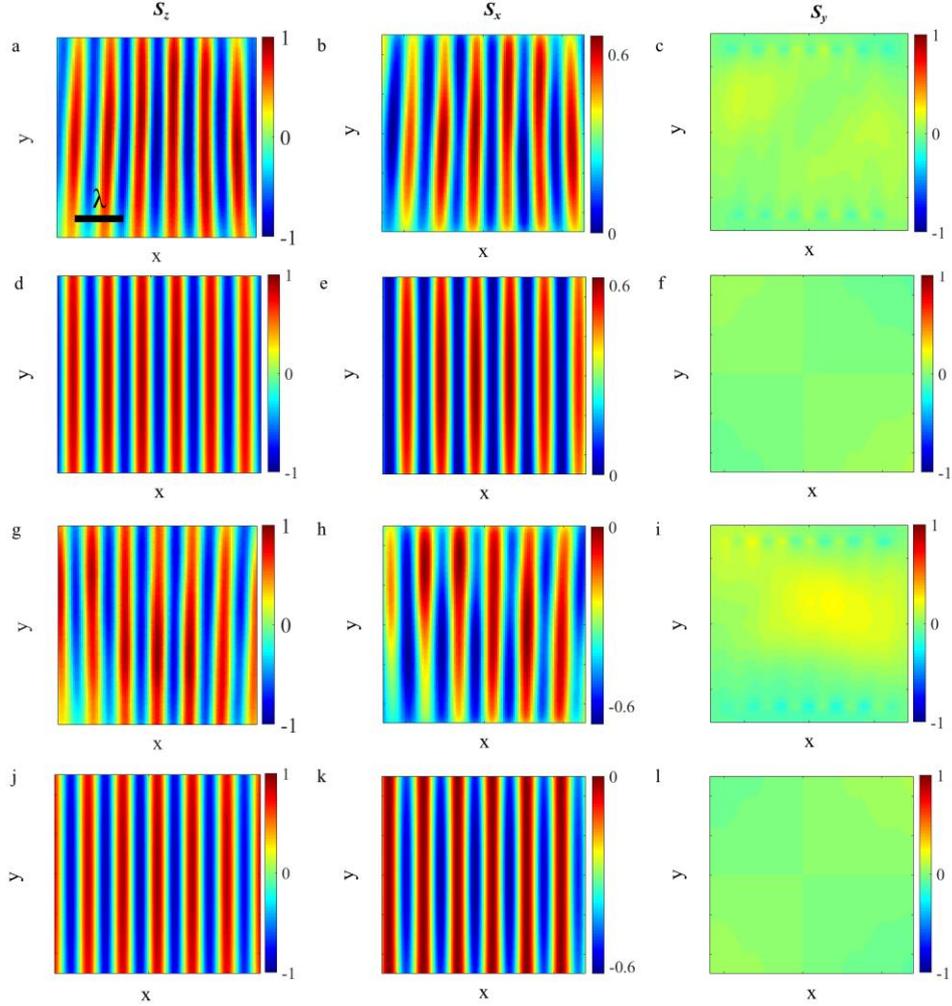

**Fig. S15 | Measured and simulated SAM components for the SPP Cosine beams. a-c** and **g-i,** experimentally obtained and **d-f** and **j-l** theoretical SAM components **a,d,g,j** $S_z$, **b,e,h,k** $S_x$ and **c,f,i,l** $S_y$ for the SPP Cosine beams propagating in **a-f** +y and **g-l** −y direction. The scale bar is the wavelength of light in vacuum.

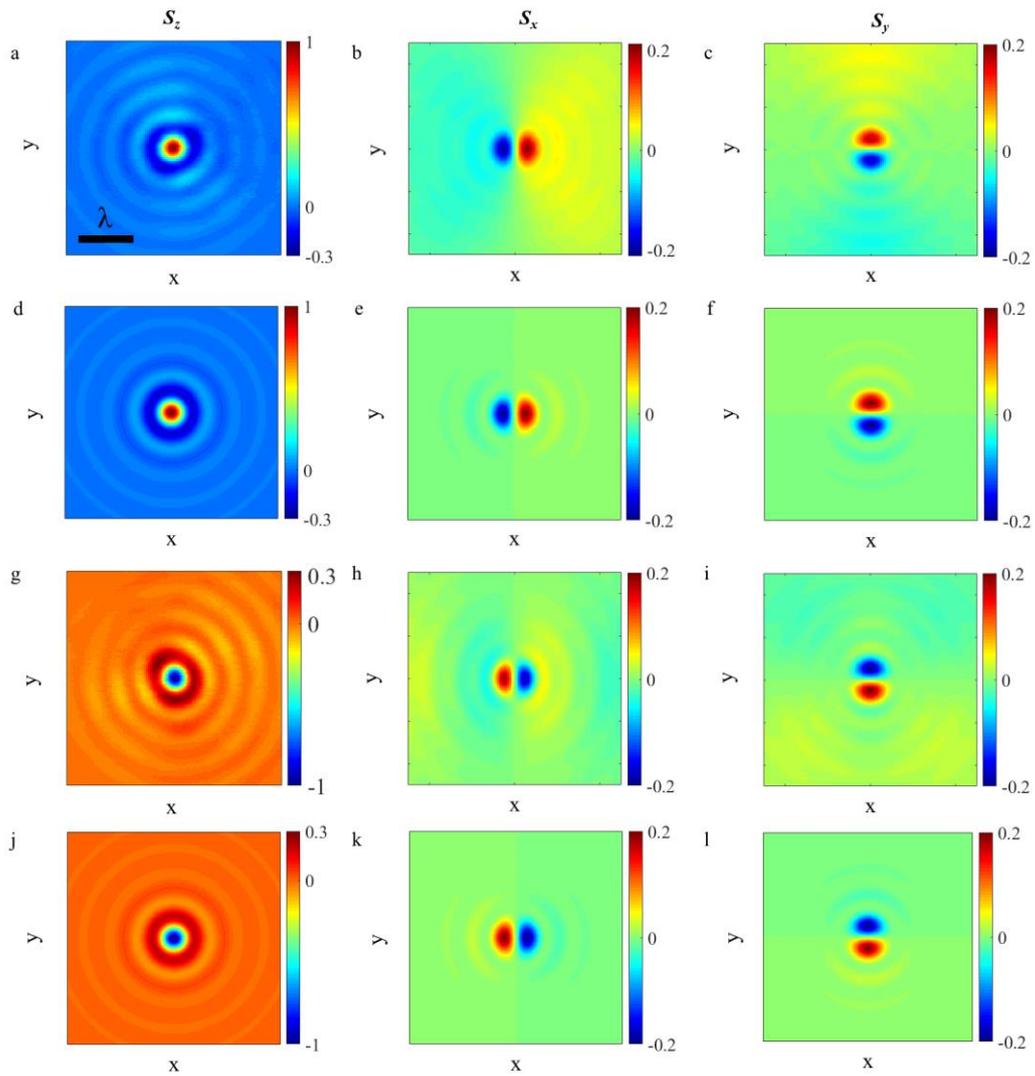

**Fig. S16 | Measured and simulated SAM components for the SPP Bessel beams. a-c** and **g-i,** experimentally obtained and **d-f** and **j-l** theoretical SAM components **a,d,g,j** $S_z$, **b,e,h,k** $S_x$ and **c,f,i,l** $S_y$ for the SPP Bessel beams with vortex topological charge **a-f** +1 and **g-l** −1. The scale bar is the wavelength of light in vacuum.

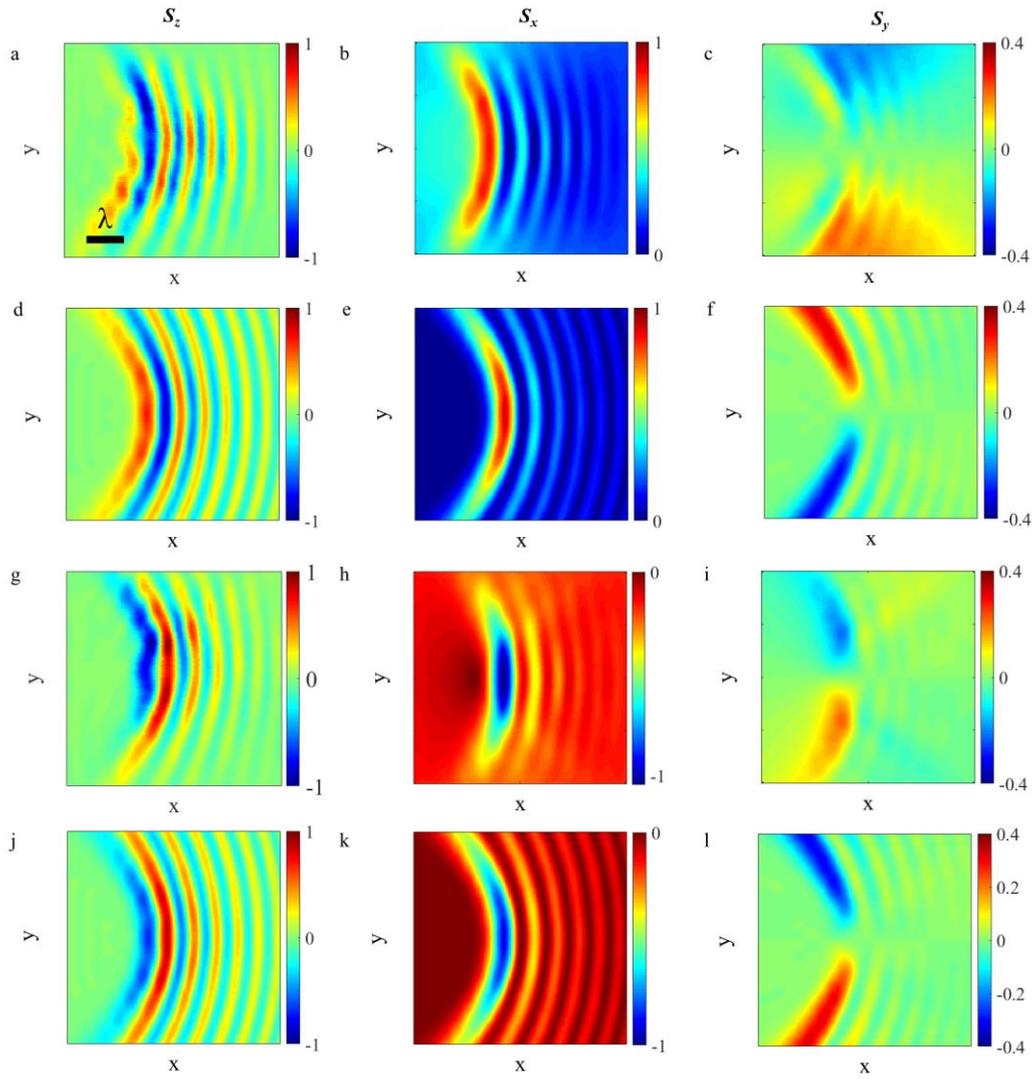

**Fig. S17 | Measured and simulated SAM components for the SPP Weber beams. a-c** and **g-i,** experimentally obtained and **d-f** and **j-l** theoretical SAM components **a,d,g,j** $S_z$, **b,e,h,k** $S_x$ and **c,f,i,l** $S_y$ for the SPP Weber beams propagating in **a-f** +**P** and **g-l** −**P** direction. The scale bar is the wavelength of light in vacuum.

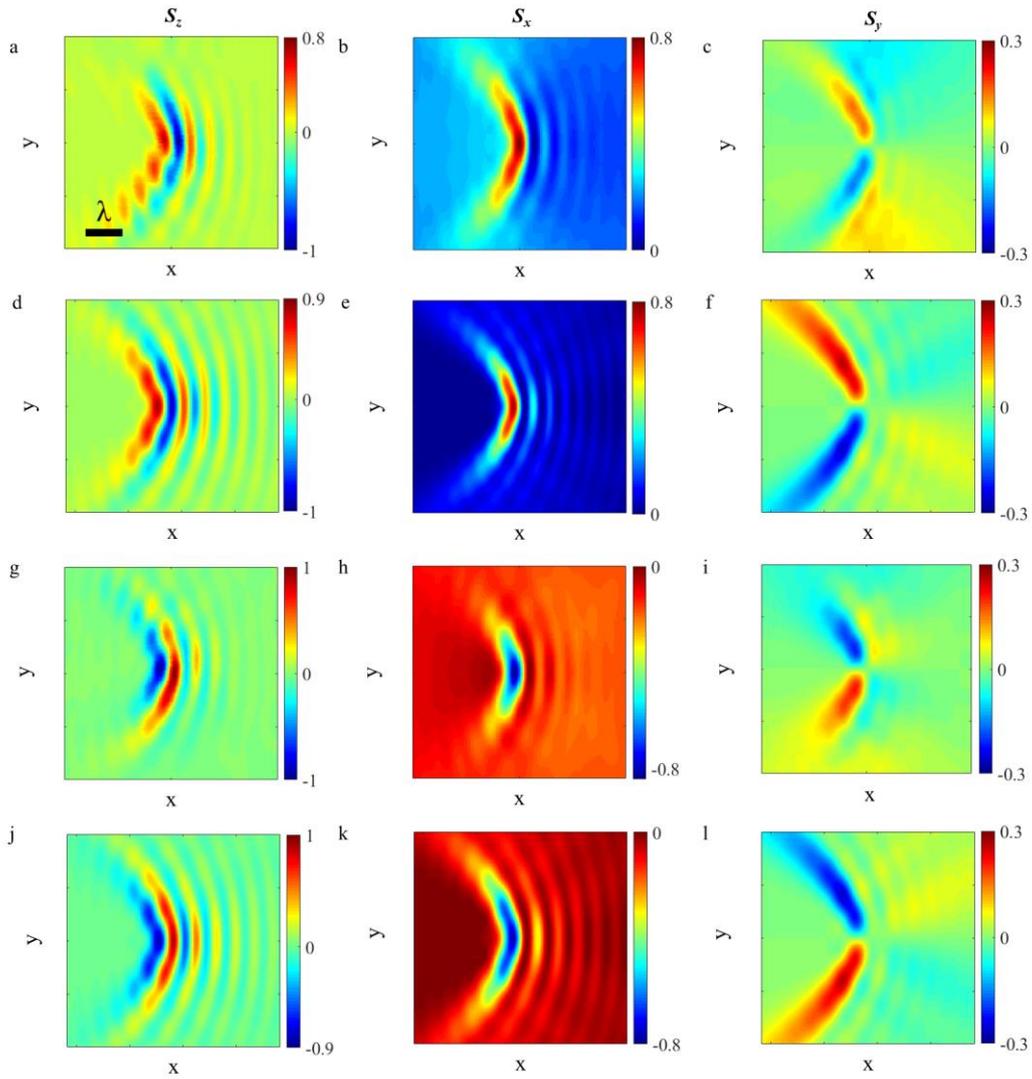

**Fig. S18 | Measured and simulated SAM components for the SPP Airy beams. a-c** and **g-i,** experimentally obtained and **d-f** and **j-l** theoretical SAM components **a,d,g,j** $S_z$, **b,e,h,k** $S_x$ and **c,f,i,l** $S_y$ for the SPP Airy beams propagating in **a-f** $+\mathbf{P}$ and **g-l** $-\mathbf{P}$ direction. The scale bar is the wavelength of light in vacuum.